\documentclass[12pt]{article}

\newcommand{\la}{\lambda}

\newcommand{\om}{\omega}

\title{ Principle of supplementarity: contextual probabilistic viewpoint  to interference, complementarity and incompatibility
}
\author{Andrei Khrennikov\\International Center for Mathematical
Modeling \\ in Physics and Cognitive Sciences,\\
University of V\"axj\"o, S-35195, Sweden\\
Email:Andrei.Khrennikov@msi.vxu.se}
\date{}
\begin{document}
\maketitle

\abstract{There is presented a contextual statistical model
of the probabilistic description of physical reality. Here contexts (complexes of physical conditions) are
considered as basic elements of reality. There is discussed the relation with  QM. We propose a
realistic analogue of Bohr's principle of complementarity. In the opposite to the Bohr's principle,
our principle has no direct relation with mutual exclusivity for observables. To distinguish our principle from the Bohr's
principle and to give better characterization, we change the terminology and speak about supplementarity,
instead of complementarity. Supplementarity is based on the interference of
probabilities. It has quantitative
expression trough a coefficient which can be easily calculated from experimental statistical data.
 We need not appeal
to the Hilbert space formalism and noncommutativity of operators representing observables. Moreover,
 in our model
there exists pairs of supplementary observables which can not be represented in the complex Hilbert space.
There are discussed applications of the principle of supplementarity outside quantum physics.}


\section{Introduction}
Since the creation of statistical mechanics, the probabilistic
approach has always played a fundamental role in physics. A crucial
step in the development of the statistical approach to physics was
made in the process of creation of quantum mechanics. It was however
soon realized by the founders of this new theory that quantum
formalism could not provide a description of physical processes for
individual systems. The understanding of this surprising fact
induced numerous debates on the possibilities of individual and
probabilistic descriptions, and on the relation between them. The
debates that followed were characterized by a wide diversity of
opinions on the origin of quantum randomness and on its relation to
classical randomness, see, e.g., Ref. 1--38 for details and
bibliography. Nowadays a rather common opinion is that the classical
probabilistic model is incompatible with the quantum one. This
opinion is based on a number of ``no-go'' theorems (von Neumann,
Kochen-Specker, Bell,...). The problem cannot however be considered
totally clarified since the {\it problem of correspondence between
quantum and classical probabilistic models} is not a purely
mathematical problem, but a physical problem. There are various
possibilities to mathematically formalize such a correspondence.
Each formalization induces its own model, which in turn usually
leads to a new proof of a ``no-go'' theorem. It should however be
stressed that these new ``no-go'' theorems suffer from the same
flaw as any other ``no-go'' theorem, in the sense that they cannot
totally eliminate the possibility of a new formalization of the {\it
quantum-classical probabilistic correspondence.}\footnote{At my talk at Beckman Institute, University of Urbana-Champain,
I was curious: Why did A. Einstein  not pay any attention to the von Neumann ``no-go´´ theorem? A. Leggett paid my attention to the fact that in the original German addition$^{(5)}$ of von Neumann's book$^{(6)}$ there was formulated not a theorem, but {\it ansatz}. So it seems that A. Einstein considered this chain of mathematical calculations as just some  arguments against hidden variables. Moreover, it might be that it was even the position of von Neumann. It might be that both Einstein and vo Neumann understood well that it is impossible to prove nonexistence of prequantum deterministic model in the form of a mathematical theorem.}  Of course, the
question of the physical adequacy should be treated separately.

In Refs. 29, 30 a new way to establish a correspondence between classical
and quantum probabilistic models was proposed. The crucial point of
this approach is that physical observables are described by a {\it
contextual probabilistic model.} The main quantum structures are
present in this model in a latent form. A quantum representation of
the contextual probabilistic model can be constructed on the basis
of {\it two specially chosen observables,} $a$ and $b.$  We call
them ``reference observables'' (e.g., the position and momentum);
only these observables are considered as representing {\it objective
properties} of physical systems (cf. the views of L. De Broglie, D.
Bohm, especially in Ref. 12 ).

This problem can be considered in a different way, as G. Mackey$^{(10)}$
did, by trying to develop a general probabilistic model ${\cal M}$
which would contain classical and quantum probabilistic models as
special cases (see also S. Gudder$^{(13)}$, L. Ballentine$^{(14)}$). The origin of the
main mathematical structures of quantum mechanics (e.g., the complex
Hilbert state space) in such a general probabilistic model should
however be clarified.

Before having a closer look at our model, it is perhaps necessary to
discuss the meaning of the term {\it contextuality}, as it can
obviously be interpreted in many different ways, see Ref. 31. The
most common meaning (especially in the literature on quantum logic$^{(33)}$)
is that the outcome for a measurement of an observable $u$ under
a contextual model is calculated using a different (albeit hidden)
measure space, depending on whether or not compatible observables
$v, w,...$ were also made in the same experiment$^{(33)}.$ We remark
that well known ``no-go'' theorems cannot be applied to such
contextual models, see  Refs. 33, 37 for details.

This approach to contextuality can be considered as a mathematical
formalization, see Ref. 33 , of {\it Bohr's measurement contextuality.}
Bohr's interpretation of quantum mechanics is in fact considered as
contextual. For N.
Bohr the word ``context'' had the meaning of a {\it ``context of a
measurement''}, see  W. De Muynck, W. De Baere, H. Marten$^{(38)},$
W. De Muynck$^{(17)}$
and A. Plotnitsky$^{(25-27)}.$ For instance, in his answer to the EPR
challenge N. Bohr pointed out that position can be determined only
in the {\it context of a position measurement.} The still unsolved
and persistent difficulty of any contextual model is that the
physical content of such theories appears in the current
understanding to be unanchored to what is obtained in any
experiment. It seems indeed difficult to explain in any satisfactory
way that a measurement on a particle $s_1$ should or could have both
a different meaning and a different associated measure space,
depending on wether or not another measurement on a second particle
$s_2$ was performed.

In our approach to the problem of correspondence between classical
and quantum probabilistic models, the term contextuality is used in
a totally different meaning. Roughly speaking our approach is
noncontextual from the conventional viewpoint$^{(33)}.$ Values
associated to the reference observables (e.g., position and
momentum) are considered as objective properties of physical
systems. These observables
are therefore not contextual in the sense of Bohr's measurement
contextuality.

The basic notion of our approach is the {\it context} -- that is, a
complex of physical conditions. Physical systems interact with a
context $C$ and in this process a statistical ensemble ${\cal S}_C$
is formed (cf. Ref. 14, 34). The notion of context is
close to the notion of preparation procedure, see e.g. Ref. 14, 33--36.
However, for any preparation procedure ${\cal
E},$ it is assumed that this procedure could be (at least in
principle) realized experimentally. We do not assume this for an
arbitrary context $C.$ Contexts are elements of physical reality which exist
independently of observers. By using the terminology of H. Atmanspacher and H. Primas$^{(32)}$ we can say that
context belong to the ontic level of description of physical reality and preparation procedure to the epistemic
level.

 Conditional (or better to say contextual)
probabilities  for reference observables, ${\bf P}(a=y/C), {\bf
P}(b=x/C),$ are used to represent the context $C$ by a complex
probability amplitude $\psi_C.$ This amplitude is in fact encoded in
a {\it generalization of the formula of total probability describing
the interference of probabilities.} Note that interferences of
probabilities can thus be obtained in a classical probabilistic
framework (i.e., without the need of the Hilbert space formalism),
an observation which was actually the starting point of our
considerations. As a result, we found that the quantum probabilistic
model can be considered as a Hilbert space projection of the
classical contextual probabilistic model. This projection is based
on two fixed ``reference observables'' $a$ and $b$ which play a
fundamental role and determine the correspondence between classical
prequantum model and quantum model.

Our approach is thus based on two cornerstones:

a) {\it contextuality of probabilities};

b) {\it the use of two fixed (incompatible) physical observables in
order to represent the classical contextual probabilistic model in
the complex Hilbert space.}

We would like to note that the conventional quantum representation
is the image of a very special class of contexts ${\cal C}^{\rm
tr}$, that is of contexts producing the usual trigonometric
interference, while other contexts producing so called hyperbolic
interference$^{(29, 30)}$ are also possible. These contexts cannot be
represented in a complex Hilbert space but in a so called hyperbolic
Hilbert space$^{(29, 30)}$ instead -- a module over a two dimensional
Clifford algebra. We will consider however in this paper only contexts having the
conventional quantum representation in a complex Hilbert space.

Our contextual statistical approach is realistic. Therefore one may
wonder how the {\it principle of complementarity} should be
interpreted or perhaps modified in such an approach. It is precisely
the purpose of this paper to study this question.

In a realistic approach one cannot just borrow Bohr's notion of
complementarity. The main problem is that of ``mutual
exclusivity'' which was considered by N. Bohr as  the main feature
of complementarity. We recall the following formulation which was
presented in Ref. 1 (vol. 2, p. 40) and was probably Bohr's most
refined formulation of what he meant by complementary measurement
situations:

\medskip

{\it Evidence obtained under different experimental conditions [e.g.
those of the position vs. the momentum measurement] cannot be
comprehended within a single picture, but must be regarded as
[mutually exclusive and] complementary in the sense that only the
totality of the [observable] phenomena exhausts the possible
information about the [quantum] objects [themselves].}

\medskip

An extended analysis of Bohr's views to complementarity can be
found in works of A. Plotnitsky$^{(25-27)}$.

Let us consider a realistic model; let $a$ and $b$ be two
observables. Then, for any physical system $\omega$, the values
$a(\omega)$ and $b(\omega)$ are simultaneously well defined (they
``coexist''). In such a situation it is rather unappropriate to
talk about the mutual exclusivity of the $a$-property and
the $b$-property for a system $\omega.$ Bohr's principle of
complementarity should therefore be modified to retain its part
related to completion (in the sense of addition of new information)
and to exclude its part related to mutual exclusivity. Hence, in
order to distinguish our contextual statistical realistic
complementarity from Bohr's complementarity, we propose to change
the terminology and use the term {\it supplementarity}, instead of
{\it complementarity.}\footnote{We understand that the change of
terminology, especially for such a fundamental principle of nature,
is a very risky choice. It might be better to use a less radical
terminology, such as V\"axj\"o principle of complementarity for
instance. However, since for most physicists complementarity is
rigidly associated with mutual exclusivity, we believe we have no
other possibility than to introduce a new terminology.

Bohr's principle of complementarity was not formulated in
mathematical terms, but as a general philosophic principle. There
are various ways to understand (or perhaps misunderstand) this
principle. The crucial point is the interpretation of ``mutual
exlusivity''. Here the difference between mathematical
{\it{variables}} and physical {\it{properties}}, as well as their
relationship with {\it{observables}}, should be taken into account,
see section 7. We emphasize that the direct comparation of the Bohr's principle of complementarity
and our principle of supplementarity is really impossible. For N. Bohr mutual exclusivity was  related not to objective properties of quantum systems, but to measurement contexts.}

In our approach the principle of complementarity-supplementarity is
formulated in mathematical terms. There exist pairs of observables
$a$ and $b$ such that $b/a$-conditioning or $a/b$-conditioning
produces supplementary information. For such observables
probabilities interfere producing nontrivial disturbances of the
classical formula of total probability.
We consider this feature of the contextual statistical realistic
model as {\it the principle of supplementarity}, or to put it in
other words, as the {\it probabilistic principle of
complementarity.}

We recall that in
the  case of two dichotomous random variables  $a=\alpha_1,
\alpha_2$ and $b=\beta_1, \beta_2$ the classical formula of total proabbility, see e.g. [39], pp. 25, 77, has the
form:
$$
{\bf P}(b=\beta_i)= {\bf P}(a=\alpha_1) {\bf
P}(b=\beta_i/a=\alpha_1)+ {\bf P}(a=\alpha_2){\bf
P}(b=\beta_i/a=\alpha_2).
$$

As was already emphasized, in the opposition to Bohr's
complementarity, supplementarity has no direct relation with mutual
exclusivity (however, see section 7 for more details). Moreover,
supplementary observables $a$ and $b$ need not present a complete
set of data in a context $C.$ In our approach, representations of
reality (e.g., physical reality) based on pairs of supplementary
observables $a$ and $b$ are in general not complete. Pairs of
supplementary observables produce very rough (statistical) images of
the underlying reality (cf. with fuzzy-viewpoint to Quantum
Mechanics, see, e.g.,  Busch, Grabowski and Lahti$^{(34)}$,   Gudder, O. Pulmanova$^{(28)}$). In
particular, in our approach the quantum complex Hilbert space is
just a projection of contextual prequantum space -- {\it
prespace}$^{(29, 30)}$.

Another difference between our supplementarity and Bohr's
complementarity is that we characterize supplementarity by studying
interferences of probabilities themselves. This gives the
possibility to introduce a simple quantitative measure of
supplementarity as a coefficient of interference $\lambda$, which we
call the {\it coefficient of supplementarity.} The presence of this
quantitative measure shifts discussion from the purely philosophic
framework to a physical and mathematical frameworks.\footnote{It is
noteworthy that our statistical measure of supplementarity $\lambda$
is not based on the Hilbert space formalism, see sections 3,4.
Hence, one need not appeal from the very beginning to the abstract
characterization based on {\it noncommutativity} of operators
representing physical observables. It is only under special
conditions that supplementary observables can be represented by
noncommutative operators in the complex Hilbert space, see Refs.
29, 30 for details. Note also that the coefficient of supplementarity
$\lambda$ can be calculated directly on the basis of experimental
statistical  data. Since we start directly with statistical data,
and since the conventional Hilbert space formalism is considered as
a secondary mathematical description, there is no necessity  to
relate our formalism with waves features or superposition of
individual states. This gives the possibility to apply our formalism
(and in particular the notion of supplementarity) outside the
quantum domain, see, e.g., Refs. 40--42.}

Our model is fundamentally contextual, and supplementarity is
consequently also contextual. The coefficient $\lambda$ depends on a
context $C$ under which observables are measured: $\lambda=
\lambda_C.$ It is therefore meaningless to discuss supplementarity
of observables $a$ and $b$ without relation to the concrete complex
of physical conditions $C$ under which these observables are
measured. Observables can indeed be supplementarity under one
context and nonsupplementarity under another one.

Since the V\"axj\"o model of reality is a statistical model, it
depends on the choice of a probability model. In this paper we use
the frequency probabilistic model (see R. von Mises$^{(43)}$, and
also Ref. 44 on this model and its applications in quantum physics).
This model is essentially more general than the conventional
Kolmogorov model$^{(45)}$. Note that other probabilistic models can
be used, e.g., the Cox model with conditional
probabilities$^{(46)}$, see L. Ballentine$^{(14)}$ on application of
this model to Quantum Mechanics.

We remark that our principle of supplementarity is based on rather
extended probabilistic considerations developed in sections 3-5. A
reader not too keen on mathematical details is invited to go through
those probabilistic sections quickly and pay more attention to
sections 6--8.

There are three appendixes, section 9 (Bell's inequality) and sections 10, 11 on mathematical features
of the contextual statistical model (relations between Kolmogorovness, compatibility and complementarity).
The latter two sections are essentially mathematical. They may be interesting for people working on mathematical problems of quantum probability and incompatibility.

\section{Contextual statistical realistic model}
From the very beginning it should be emphasized that the purpose of
this section is not to provide a new interpretation of Quantum
Mechanics. A general statistical model for observables based on the
contextual viewpoint to probability will merely be presented. It
will be shown that classical as well as quantum probabilistic models
can be obtained as a particular cases of our general contextual
model, the {\it{V\"axj\"o model}}.
    This model is not reduced to the conventional, classical and quantum
    models. In particular, it contains a new statistical model: a model with
hyperbolic $cosh$-interference (that induces  "hyperbolic quantum
mechanics").\footnote{It should however be noted that hyperbolic model remains a purely
mathematical model, in contrast to those of classical physics and
quantum mechanics, as it does not relate to any known physics.}

Realism is one of the main distinguishing features of
the V\"axj\"o model since it is always possible manipulate
objective properties, despite the presence of such essentially
quantum effects as, e.g., the interference of probabilities.

As George W. Mackey pointed out in Ref. 10, probabilities cannot be
considered as abstract quantities defined outside any reference to a
concrete complex of physical conditions $C.$ All probabilities are
conditional or better to say contextual.\footnote{The notion of
conditional probability is typically used for events:
${\bf P}(A/B)$ is the probability that the event $A$ occurs
under the condition that the event $B$ has occurred.} We remark
that the same point of view can be found in the works of A. N.
Kolmogorov$^{(45)}$ and R. von Mises$^{(43)}$. However, it seems that Mackey's
book was the first thorough presentation of a program of conditional
probabilistic description of measurements, both in classical and
quantum physics. G. Mackey did a lot to unify classical and quantum
probabilistic description and, in particular, demystify quantum
probability. One crucial step is however missing in Mackey's work.
In his book$^{(10)},$ Mackey introduced the quantum probabilistic model
(based on the complex Hilbert space) by means of a special axiom
(Axiom 7, p. 71, in Ref. 10) that looked rather artificial in his
general conditional probabilistic framework. The impossibility to
derive the quantum probabilistic model from ``natural axioms''
(which are not based on such a quantum structure as the complex
Hilbert space) is clearly the main disadvantage of Mackey's
approach.

In our contextual probabilistic approach, structures specifically
belonging to the realm of quantum mechanics (e.g., the interference
of probabilities, the complex probabilistic amplitudes, Bohr's
rule, or the representation of some observables by noncommutative
operators) are derived on the basis of two natural axioms (that is,
natural from the point of view of classical probabilistic
axiomatics).

Mackey's model is based on a system of eight axioms, when our own
model requires only two axioms. Let us briefly mention the content
of Mackey first axioms. The first four axioms concern conditional
structure of probabilities, that is, they can be considered as
axioms of a classical probabilistic model. The fifth and sixth
axioms are of a logical nature
    (about questions).
We reproduce below Mackey's ``quantum axiom'', and Mackey's own
comments on this axiom (see Ref. 10, pp. 71-72):

\medskip

{\bf Axiom 7} (G. Mackey) {\it The partially ordered set of all questions in quantum
mechanics is isomorphic to the partially ordered set of all closed
subsets of a separable, infinite dimensional Hilbert space.}

\medskip

{\small ``This axiom has rather a different character from Axioms 1
through 4. These all had some degree of physical naturalness and
plausibility. Axiom 7 seems entirely {\it ad.hoc.} Why do we make
it? Can we justify making it? What else might we assume? We shall
discuss these questions in turn. The first is the easiest to answer.
We make it because it ``works'', that is, it leads to a theory
which explains physical phenomena and successfully predicts the
results of experiments. It is conceivable that a quite different
assumption would do likewise but this is a possibility that no one
seems to have explored [But see recent work of Jauch, Stueckelberg,
and others at the University of Gen\`{e}ve on real and quaternionic
Hilbert spaces]. Ideally one would like to have a list of physically
plausible assumptions from which one could deduce Axiom 7.''}

Our activity can be considered as an attempt to find a list of
physically plausible assumptions from which the Hilbert space
structure can be deduced. We show that this list can consist in two
axioms (see our Axioms 1 and 2) and that these axioms can be
formulated in the same classical probabilistic manner as Mackey's
Axioms 1--4.

Another important difference between the V\"axj\"o model and
Mackey's model is that our model is not rigidly coupled to the
measure theoretic approach to probability. Many probabilistic models
can be used to mathematically define contextual probabilities. In
this paper we use the von Mises' frequency approach to probability$^{(43)}.$
This approach is essentially more general than the Kolmogorov
measure-theoretic one. In general probabilistic data generated by a
few collectives, $x, y, z, \ldots, u,$ cannot be described by a
single Kolmogorov space. There are admittedly some mathematical
difficulties in von Mises' approach, like the impossibility to
rigorously define randomness on a mathematical level. Nevertheless,
in order to define probabilities one need not necessarily to apply
von Mises' principle of randomness (which is based on a rather
confusing notion of place selection). The frequency probability can
be defined by the principle of statistical stabilization of relative
frequencies. It means that relative frequencies $\nu_N=\frac{n}{N}$
stabilize when $N \to \infty,$ i.e., $|\nu_N - \nu_M| \to 0, N, M
\to \infty.$ Hence, the limit of $\nu_N$ exists and is called the
frequency probability. The frequency theory of probability, which is
based only on the principle of statistical stabilization is free of
contradictions$^{(44)}.$

Quantum formalism gives good predictions of frequency probabilities,
as was verified in an impressive number of experiments, but it does
not contain a description of randomness. In the light of the
approach described above, the theory of quantum measurements appears
to be more about statistical stabilization of relative frequencies
than about randomness of data. It seems that in order to create a
general frequency probability theory, which would contain classical
and quantum probabilities as special cases, one could use the
frequency probabilistic model based only on the principle of
statistical stabilization.
\subsection{Contextual statistical model of observations}
A physical {\it context}  $C$ is  a complex of physical conditions.
Contexts are fundamental elements of any contextual statistical model. Thus construction of any model
${\cal M}$ should be started with fixing the collection of  contexts of this model; denote the collection of contexts
by the symbol ${\cal C}$ (so ${\cal C}$ is determined by ${\cal M}).$ In mathematical formalism ${\cal C}$ is an abstract set
(of ``labels'' of contexts). Another fundamental element of any contextual statistical model ${\cal M}$ is a set of observables ${\cal O}:$
any observable $a\in {\cal O}$ can be measured
under a complex of physical conditions $C\in {\cal C}.$ \footnote{We shall
denote observables by Latin letters, $a,b,...,$ and their values by Greek letters,
$\alpha, \beta,...$  } For an $a \in {\cal O},$ we denote the set of its possible values (``spectrum'') by the symbol
$X_a.$

We do not assume that all these observables can be measured simultaneously;
so they need not be compatible. To simplify considerations, we shall consider only discrete observables
and, moreover, all concrete investigations will be performed for {\it dichotomous observables.}

\medskip

{\bf Axiom 1:} {\it For  any observable
$a \in {\cal O},$ there are defined contexts $C_\alpha$
corresponding to $\alpha$-filtrations: if we perform a measurement of $a$ under
the complex of physical conditions $C_\alpha,$ then we obtain the value $a=\alpha$ with
probability 1. It is supposed that the set of contexts ${\cal C}$ contains filtration-contexts $C_\alpha$
for all observables $a\in {\cal O}.$}

\medskip

{\bf Axiom 2:} {\it There are defined contextual probabilities ${\bf P}(a=\alpha/C)$ for any
context $C \in {\cal C}$ and any observable $a \in {\it O}.$}

\medskip

Probabilities ${\bf P}(b=\beta/C)$ are interpreted as {\it contextual probabilities.} Especially important role will be played by probabilities:
$$
p^{b/a}(\beta/\alpha)\equiv {\bf P}(b=\beta/C_\alpha), a, b \in {\cal O}, \alpha \in X_a, \beta \in X_b,
$$
where $C_\alpha$ is the $[a=\alpha]$-filtration context.\footnote{
We prefer to call probabilities with respect to a context $C\in {\cal C}$ contextual probabilities.
Of course, it would be also possible to call them conditional, but the latter term was already used in
other approaches (e.g., Bayes-Kolmogorov).
In the opposition of the Bayes-Kolmogorov model,
the contextual probability is not probability that an event, say  $B,$ occurs under the condition that another event,
say $C,$ has been occurred. The contextual probability  ${\bf P}(b=\beta/C)$ is probability to get the
result
$b=\beta$ under the complex of physical conditions $C.$ We can say that this is the probability that the
event $B_\beta=\{ b=\beta\}$ occurs under the complex of physical conditions $C.$ Thus in our approach not event, but context should be considered as a condition; see also
Accardi$^{(15)}$, Ballentine$^{(14)}$  and De Muynck$^{(17)}.$
In particular, the contextual probability
$
p^{b/a}(\beta/\alpha)\equiv {\bf P}(b=\beta/C_\alpha),
$
is not probability that the event
$B_\beta=\{b=\beta\}$ occurs under the condition that the event
$A_\alpha=\{ a=\alpha\}$ has been occurred. To find probability $p^{b/a}(\beta/\alpha),$ it is
not sufficient to observe
the event $B_\beta$ following  the event $A_\alpha.$ It should be first verified that there is
really the complex of physical conditions $C_\alpha.$  Then there are performed measurements of
the observable $b$ under this context, see also Ballentine$^{(14)}$  and De Muynck$^{(17)}.$}

At the moment we do not fix
a definition of  probability. Depending on a choice of probability theory we can obtain
different models. For any $C\in {\cal C},$ there is defined the set of probabilities:
$$
E({\cal O}, C)= \{ {\bf P}(a=\alpha/C): a \in {\cal O}\}
$$
We complete this probabilistic data by $C_\alpha$-contextual probabilities:
$
D({\cal O}, C)= \{ {\bf P}(a=\alpha/C),{\bf P}(b=\beta/C), ...,
{\bf P}(a=\alpha/C_\beta), {\bf P}(b=\beta/C_\alpha),...:a,b,... \in {\cal O}\}.
$

We remark that $D({\cal O}, C)$ does not contain the simultaneous probability distribution of
observables ${\cal O}.$
Data $D({\cal O}, C)$ gives a probabilistic image of the context $C$ through the
system of observables ${\cal O}.$
We denote by the symbol ${\cal D}({\cal O}, {\cal C})$ the collection of probabilistic data
$D({\cal O}, C)$ for all contexts  $C\in {\cal C}.$ There is defined the map:
\begin{equation}
\label{MP}
\pi :{\cal C} \to {\cal D}({\cal O}, {\cal C}), \; \; \pi(C)= D({\cal O}, C).
\end{equation}
In general this map is not one-to-one. Thus the $\pi$-image of contextual reality is very rough:
{\it not all contexts can be distinguished with the aid of probabilistic data produced by the class
of observables ${\cal O}.$ }

{\bf Definition 2.1.} {\it A contextual  statistical model of reality is a triple
\begin{equation}
\label{VM}
M =({\cal C}, {\cal O}, {\cal D}({\cal O}, {\cal C}))
\end{equation}
where ${\cal C}$ is a set of contexts and ${\cal O}$ is a  set of observables
which satisfy to axioms 1,2, and ${\cal D}({\cal O}, {\cal C})$ is probabilistic data
about contexts ${\cal C}$ obtained with the aid of observables ${\cal O}.$}

We call observables belonging the set ${\cal O}\equiv {\cal O}(M)$ {\it reference of observables.}
Inside of a model $M$  observables  belonging ${\cal O}$ give the only possible references
about a context $C\in {\cal C}.$ Our general model can (but, in principle,
need not) be completed by some interpretation of reference observables $a\in {\cal O}.$
By the V\"axj\"o interpretation reference observables are interpreted as {\it properties of contexts.}

\medskip

{\bf Realistic interpretation of observables:}
``If an observation of $a$ under a complex of physical
conditions $C \in {\cal C}$ gives the result $a=\alpha,$  then this value is interpreted as
the objective property of the context $C$ (at the moment of the observation).''

\medskip

{\bf Remark 2.2.} (Number of reference observables) In both most important physical models --
in classical and quantum models --
the set ${\cal O}$ of reference observables consists of {\bf two observables:}
{\it position and momentum (or energy).}
I think that this number ``two'' of reference observables plays the crucial role
(at least in the quantum model).

\subsection{Frequency description of probability distributions}
By taking into account Remark 2.2,
we consider a set of reference
observables ${\cal O}= \{ a, b \}$ consisting of two observables $a$ and $b.$
We denotes the sets of values (``spectra'') of the reference observables by symbols $X_a$ and $X_b,$
respectively.

Let $C$ be some context. In a series of observations of $b$ (which can be infinite in a mathematical model)
we obtain a sequence of values of $b:$
\begin{equation}
\label{KOL1}
x\equiv x(b/C) = (x_1, x_2,..., x_N,...), \;\; x_j\in X_b.
\end{equation}
In a series of observations of $a$ we obtain a sequence of values of $a:$
\begin{equation}
\label{KOL2}
y\equiv y(a/C) = (y_1, y_2,..., y_N,...), \;\; y_j\in X_a.
\end{equation}
We suppose that the {\it principle of the statistical stabilization} for relative frequencies$^{(43, 44)}$
holds true. This means that the frequency probabilities
are well defined:
\begin{equation}
\label{KOL3}
p^b(\beta) \equiv {\bf P}_x( b=\beta)= \lim_{N\to \infty} \nu_N(\beta; x), \;\; \beta \in X_b;
\end{equation}
\begin{equation}
\label{KOL3a}
p^a(\alpha) \equiv {\bf P}_y( a=\alpha)= \lim_{N\to \infty} \nu_N(\alpha; y), \;\; \alpha\in X_a.
\end{equation}
Here $\nu_N(\beta; x)$ and $ \nu_N(\alpha; y)$ are frequencies of observations of values
$b=\beta$ and $a=\alpha,$ respectively (under the complex of conditions $C).$

As was remarked, R. von Mises considered
in his theory two principles: a) {\it the principle of the statistical stabilization for relative frequencies};
 b){\it the principle of randomness.} A sequence of observations for which
both principle hold was called a {\it collective.}  An analog of von Mises' theory for sequences of observations
which satisfy the principle of statistical stabilization (so relative frequencies converge to limit-probabilities, but
these limits need not be invariant with respect to von Mises place selections)
was developed in Ref. 44; we call such sequences $S$-{\it sequences.}
Everywhere in this paper it will be assumed that {\it sequences of observations are $S$-sequences},
cf. Ref. 44.

Let $C_{\alpha},  \alpha\in X_a,$  be contexts  corresponding
to  $\alpha$-filtrations, see Axiom 1.
By observation of $b$ under the context $C_\alpha$ we obtain a sequence:
\begin{equation}
\label{KOL4}
x^{\alpha} \equiv x(b/C_\alpha) = (x_1, x_2,..., x_{N},...), \;\; x_j \in X_b.
\end{equation}
It is also assumed that for  sequences of observations  $x^{\alpha}, \alpha\in X_a,$
the principle of statistical stabilization for relative frequencies
holds true and the frequency probabilities are well defined:
\begin{equation}
\label{KOL5}
p^{b/a}(\beta/\alpha) \equiv {\bf P}_{x^{\alpha}}(b=\beta)= \lim_{N \to \infty} \nu_{N}(\beta; x^{\alpha}), \;\;
\beta \in X_b.
\end{equation}
Here $\nu_N(\beta; x^\alpha), \alpha\in X_a,$  are frequencies of observations of value
$b=\beta$ under the complex of conditions $C_\alpha.$ We obtain
probability distributions:
\begin{equation}
\label{KKK4}
{\bf P}_x(\beta), \;\; {\bf P}_y (\alpha), \;
{\bf P}_{x^{\alpha}}(\beta),\;\;\alpha\in X_a, \beta \in X_b.
\end{equation}
We can repeat all previous considerations by changing $b/a$-conditioning to  $a/b$-conditioning.
We consider
contexts $C_\beta, \beta \in X_b,$ corresponding to selections with respect to values of the
observable $b$ and the
corresponding collectives $y^{\beta}\equiv y(a/C_\beta)$ induced by   observations of $a$ in
 contexts $C_\beta.$
There can be defined probabilities $p^{a/b}(\alpha/\beta)\equiv {\bf P}_{y^{\beta}}(\alpha).$
Combining these
data with data (\ref{KKK4}) we obtain
$$
D({\cal O}, C)= \{ p^a(\alpha), p^b(\beta), p^{b/a}(\beta/\alpha), p^{a/b}(\alpha/\beta): \alpha\in X_a, \beta \in X_b\}
$$

\subsection{Systems, ensemble representation}

 We now complete the
contextual statistical model
by considering systems $\omega$ (e.g., physical or cognitive, or social,..), (Cf. Ballentine$^{(14)}).$
In our approach systems as well as contexts are considered as {\it elements of realty. }
In our model a context $C \in {\cal C}$ is represented  by an ensemble $S_C$ of systems which have
been interacted  with $C.$ For such systems we shall use notation:
$
\omega \hookleftarrow C
$
The set of all (e.g., physical or cognitive, or social)
systems which are used to represent all contexts $C\in {\cal C}$ is denoted by the symbol
$\Omega\equiv \Omega({\cal C}).$
Thus we have a map:
\begin{equation}
\label{VMM}
C \to S_C=\{ \omega\in \Omega:  \omega \hookleftarrow C \}.
\end{equation}
This is the ensemble representation of contexts. We set
$$
{\cal S}\equiv {\cal S}({\cal C})=\{S: S=S_C, C \in {\cal C}\}.
$$
This is the collection of all ensembles representing contexts belonging to ${\cal C}.$
The ensemble representation of contexts is given by the map (\ref{VMM})
$$
I: {\cal C} \to {\cal S}
$$
Reference observables ${\cal O}$ are now interpreted as observables on systems $\omega\in \Omega.$
In our approach it is not forbidden to interpret the values of the {\it reference observables} as objective properties
of systems. These objective properties coexist in nature and they can be related to individual systems $\omega \in \Omega.$ However, the probabilistic description is possible only with respect to a fixed context $C.$ Noncontextual probabilities have no meaning. So values $a(\omega)$ and $b(\omega)$ coexist for a single system $\omega\in \Omega,$
but noncontextual (``absolute'')  probabilities ${\bf P}(\omega \in \Omega: a(\omega)=y), ...$ are not defined.
\footnote{Thus, instead of mutual exclusivity of observables (cf. Bohr's principle of complementarity), we consider contextuality of probabilities and ``supplementarity'' of the reference observables (in the sense that they give supplementary statistical
information about contexts). Oppositely to the very common opinion, such models (with realistic observables)
can have nontrivial quantum-like representations (in complex and hyperbolic Hilbert spaces)
which are based on the formula of total probability with interference terms.}

{\bf Definition 2.2.} {\it The ensemble representation of a contextual  statistical model
$M =({\cal C}, {\cal O}, {\cal D}({\cal O}, {\cal C}))$ is a triple
 \begin{equation}
 \label{VM1}
 S(M) =({\cal S}, {\cal O}, {\cal D}({\cal O}, {\cal C}))
 \end{equation}
 where ${\cal S}$ is a set of ensembles representing contexts ${\cal C}$,
 ${\cal O}$ is a  set of observables, and ${\cal D}({\cal O}, {\cal C})$ is probabilistic data
 about ensembles ${\cal S}$ obtained with the aid of observables ${\cal O}.$}

\section{Formula of total probability and measures of supplementarity}
Let $M =({\cal C}, {\cal O}, {\cal D}({\cal O}, {\cal C}))$ be a model in which
${\cal O}= \{a, b\}$ and $a, b$ are dichotomous observables. Let $C \in {\cal C}.$
In general there are no reasons to assume
that all probability distributions in $D(a, b, C)$ should be described by a single Kolmogorov
probability space (absolute Kolmogorov space) ${\cal P}= (\Omega, {\it F}, {\bf P}).$ Thus the classical
(Kolmogorovian) formula of total probability:
\begin{equation}
\label{TRP}
{\bf P}(b=\beta) = \sum_\alpha {\bf P}(a=\alpha){\bf P}(b=\beta/a=\alpha).
\end{equation}
can be violated. We do not have such a formula in the contextual
frequency approach, where the conditional probabilities ${\bf
P}(b=\beta/a=\alpha)$ are defined as contextual probabilities ${\bf
P}(b=\beta/C_\alpha).$\footnote{We recall that this is the {\it
premeasurement} conditioning. The complex of physical conditions
$C_\alpha,$ corresponding to $[a=\alpha]$-selection,  is fixed
before the $b$-measurement.} In this approach everything is
absolutely clear from the very beginning: there are 6 different
$S$-sequences (or collectives), see section 2.2:
$$
x=x(b/C), y=y(a/C), x^\alpha=x(b/C_\alpha), y^\beta=y(a/C_\beta),
$$
where $\alpha=\alpha_1, \alpha_2, \beta= \beta_1, \beta_2.$
In the opposition
to the Kolmogorov approach, in the contextual frequency approach we have no chance to speculate about a
single probability.
In the contextual frequency approach  in general we have:
\begin{equation}
\label{KOL6}
\delta(\beta/ a, C) = {\bf P}_x(\beta) - \sum_\alpha {\bf P}_y(\alpha){\bf P}_{x^{\alpha}}(\beta) \not=0
\end{equation}
On the other hand, as was mentioned, in the Kolmogorov model
we have:
\begin{equation}
\label{KOL7}
\delta(\beta/ a, C)=0.
\end{equation}

Hence, in the Kolmogorov model
by using the Bayesian sum of  the  conditional probabilities ${\bf P} (b=\beta/a=\alpha)$ we find
nothing new, but the unconditional probability for $b=\beta:$
$$
{\bf P}(b=\beta) = \sum_\alpha {\bf P}(a=\alpha){\bf P}(b=\beta/a=\alpha).
$$
Therefore in these models by obtaining the value $b=\beta$ in a series of observations
under the condition $a = \alpha$  we do not obtain
new probabilistic information. However, in the contextual approach we obtain new information
via conditional observations, see (\ref{KOL6}).
Hence conditional observations give us {\it supplementary information} which is not contained
in statistical data for unconditional observations.
\medskip

{\bf Definition 3.1.} {\it The quantity $\delta(\beta/ a, C)$ is said to be
a probabilistic measure of $b/a$-supplementarity in the context $C.$}

\medskip

We can rerwite the equality (\ref{KOL6}) in the form which is similar to the classical formula
of total probability:
\begin{equation}
\label{KOL8}
{\bf P}_x(\beta) = \sum_\alpha {\bf P}_y(\alpha) {\bf P}_{x^{\alpha}}(\beta) + \delta(\beta/ a, C),
\end{equation}
or by using shorter notations:
\begin{equation}
\label{KOL9}
p^b(\beta)= \sum_\alpha p^a(\alpha) p^{b/a}(\beta/\alpha) + \delta(\beta/ a, C).
\end{equation}
This formula has the same structure as the quantum formula of total probability:

\medskip

[classical part] + additional term.

\medskip

To write the additional term in the same form as in the quantum representation of statistical
data, we perform the normalization
of the probabilistic measure of supplementarity  by the square root of the product of all
probabilities:

\medskip

\begin{equation}
\label{KOL10}
\lambda(\beta/ a, C)= \frac{\delta(\beta/ a, C)}{2\sqrt{\prod_\alpha p^a(\alpha) p^{b/a}(\beta/\alpha)}} .
\end{equation}

\medskip

{\it The coefficient $\lambda(\beta/ a, C)$ also will be called the probabilistic measure of supplementarity.}

\medskip

Of course, it would be better to call $\lambda$ the coefficient of complementarity, but the latter
terminology was already reserved by N. Bohr.
By using this coefficient we rewrite (\ref{KOL9}) in the quantum-like form:
\begin{equation}
\label{KOL11}
p^b(\beta) = \sum_\alpha p^a(\alpha) p^{b/a}(\beta/\alpha) +
2 \lambda(\beta/ a, C) \sqrt{\prod_\alpha p^a(\alpha) p^{b/a}(\beta/\alpha)}.
\end{equation}

The coefficient $\lambda(\beta/ a, C)$ is well defined only in the case when
all probabilities $p^a(\alpha), p^{b/a}(\beta/\alpha)$ are strictly positive. We consider the matrix
$$
{\bf P}^{b/a}= ( p^{b/a}(\beta/\alpha) )
$$
Traditionally this matrix  is called the matrix of transition probabilities.
In our approach $p^{b/a}(\beta/\alpha)\equiv
{\bf P}_{x^{\alpha}}(b= \beta)$ is the probability to obtain the value $b=\beta$ for the $S$-sequence (collective)
$x^{\alpha}.$ Thus in general  we need not speak about states of physical systems and interpret
$p^{b/a}(\beta/\alpha)$ as the probability of the transition from the state $\alpha$ to the state
$\beta.$ We remark that the matrix ${\bf P}^{b/a}$ is always {\it stochastic}:
\begin{equation}
\label{SS}
\sum_\beta p^{b/a}(\beta/\alpha) =1
\end{equation}
for any $\alpha\in X_a,$ because for any $S$-sequence (or collective) $x^{\alpha}:$
$$
\sum_\beta {\bf P}_{x^{\alpha}}(b= \beta)=1 .
$$

We defined a {\it nondegenerate} $S$-sequence (or collective) $y$ as such that
$$
p^a(\alpha)\equiv {\bf P}_y(\alpha)\not= 0 \;\; \mbox{for all}\;\; \alpha.
$$
A context $C$ is said to be $a$-nondegenerate
($b$-nondegenerate) if the corresponding
$S$-sequence (or collective) $y\equiv y(a/C)$ ($x\equiv x(b/C))$ is nondegenerate.
We remark that the contexts $C_\alpha$ ($S$-sequences or collectives) $x^\alpha)$
are $b$-nondegenerate iff
\begin{equation}
\label{TIR}
p^{b/a}(\beta/\alpha)\not=0.
\end{equation}

The representation (\ref{KOL11}) is the basis of transition  to
a (complex or hyperbolic) Hilbert space representation of probabilistic data $D(a, b, C).$
The representation (\ref{KOL11}) can be used only for  nondegenerate contexts $C$ and  $C_\alpha.$

We can repeat all previous considerations by
changing $b/a$-conditioning to  $a/b$-conditioning. We consider
contexts $C_\beta$ corresponding to selections with respect to values of the observable $b$ and the
corresponding $S$-sequences (or collectives) $y^{\beta}\equiv y(a/C_\beta).$
There can be defined  probabilistic measures of supplementarity $\delta(\alpha/b, C)$ and
$\lambda(\alpha/b, C), \alpha \in X_a.$ We remark that the contexts $C_\beta$
($S$-sequences or collectives $y^\beta)$ are $a$-nondegenerate iff
\begin{equation}
\label{TIRS}
p^{a/b}(\alpha/\beta) \not= 0.
\end{equation}

For  nondegenerate contexts $C$ and $C_\beta$ we have:
\begin{equation}
\label{KaF}
p^a(\alpha) = \sum_\beta p^b(\beta) p^{a/b}(\alpha/\beta) +
2 \lambda(\alpha/b, C) \sqrt{\prod_\beta p^b(\beta)  p^{a/b}(\alpha/\beta)}.
\end{equation}

{\bf Definition 3.2.} {\it Observables $a$ and $b$ are called (statistically) nondegenerate if (\ref{TIR}) and
(\ref{TIRS}) hold.}

{\bf Theorem 3.1.} {\it Let reference observables $a$ and $b$  be nondegenerate  and  let a context $C\in {\cal C}$
be both $a$ and $b$-nondegenerate. Then quantum-like formulas of total probability (\ref{KOL11})
and (\ref{KaF}) hold true.}

In Ref. 1, 22  Theorem 3.1 was proved by using a long series of calculations with relative frequencies; the proof
 presented in this paper is really straightforward: supplementarity
implies the violation of the classical formula of total probability and the
perturbation term can be represented in the quantum-like form. As was shown in
Refs. 1-3, if $\vert \lambda \vert\leq 1,$ then we get $\cos$-interference (see section ??);
if $\vert \lambda \vert\ > 1,$ then we get $\cosh$-interference.

\section{Supplementary physical observables}

{\bf Definition 4.1.} {\it  Reference observables $a$ and $b$ are called
$b/a$-supplementary in a context $C$ if}
\begin{equation}
\label{C1}
\delta(\beta/a, C) \ne 0 \; \mbox{for some}\; \beta \in X_b.
\end{equation}

{\bf Lemma 4.1.} {\it For any context $C\in {\cal C},$ we have:}
\begin{equation}
\label{CA1}
\sum_{\beta\in X_b} \delta(\beta/a, C) = 0
\end{equation}

{\bf Proof.}  We have
$$
1= \sum_{\beta\in X_b} p^b(\beta) =  \sum_{\beta\in X_b}\sum_{\alpha\in X_a}  p^a(\alpha) p^{b/a}(\beta/\alpha) +
\sum_{\beta\in X_b} \delta(\beta/a, C).
$$
Since ${\bf P}^{b/a}$ is always a stochastic matrix, we have for any $\alpha \in X_a:$
$$
\sum_{\beta\in X_b} p^{b/a}(\beta/\alpha)=1.
$$
By using that $\sum_{\alpha\in X_a}  p^a(\alpha) =1$ we obtain (\ref{CA1}).

\medskip
We pay attention to the fact that by Lemma 5.1 the coefficient $\delta(\beta_1/a, C)= 0$
iff $\delta(\beta_2/a, C)=0.$ Thus $b/a$-supplementarity is equivalent to the condition
$\delta(\beta/a, C)\not= 0$ both for  $\beta_1$ and $\beta_2.$

Reference observables $a$ and $b$ are called supplementary in a context $C$ if they
are   $b/a$ or $a/b$ supplementary:
\begin{equation}
\label{C2}
\delta(\beta/a, C) \ne 0 \;\mbox{or}\; \delta(\alpha/b, C) \ne 0 \; \mbox{for some}\; \beta \in X_b,
\alpha \in X_a.
\end{equation}

By Lemma 4.1 observables are supplementary iff the coefficients $\delta(\beta/a,C) =0$ and
$\delta(\alpha/b, C)=0$ for all $\beta \in X_b,\alpha \in X_a.$

Let us consider a contextual model $M$ with the set of contexts ${\cal C}.$
Observables $a$ and $b$ are said to be supplementary in the model $M$
if there exists $C\in {\cal C}$ such that they are supplementary in the context $C.$

Reference observables $a$ and $b$ are said to be nonsupplementary
in the context $C$ if they are neither $b/a$ nor $a/b$-supplementary:
\begin{equation}
\label{C3}
\delta(\beta/a, C) = 0 \;\mbox{and}\; \delta(\alpha/b, C) = 0\; \mbox{for all}\;\beta \in X_b,
\alpha \in X_a.
\end{equation}
Thus in the case of $b/a$-supplementarity we have (for $\beta\in X_b):$
\begin{equation}
\label{bC1}
p^b(\beta) \ne \sum_\alpha p^a(\alpha) p^{b/a}(\beta/\alpha);
\end{equation}
in the case of  $a/b$-supplementarity we have (for $\alpha\in X_a)$:
\begin{equation}
\label{bC2}
p^a(\alpha) \ne \sum_\beta p^b(\beta) p^{a/b}(\alpha/\beta);
\end{equation}
in the case of supplementarity we have (\ref{bC1}) or (\ref{bC2}). In the case of nonsupplementarity
we have both representations:
\begin{equation}
\label{bC3}
p^b(\beta)=\sum_\alpha p^a(\alpha) p^{b/a}(\beta/\alpha), \; \beta\in X_b,
\end{equation}
\begin{equation}
\label{bC4}
p^a(\alpha)=\sum_\beta p^b(\beta) p^{a/b}(\alpha/\beta),\; \alpha\in X_a.
\end{equation}

\section{The principle of supplementarity}
Our principle can be formulated in the following way:
{\it{ There exist physical observables, say $a$ and $b$, such that for some context $C$ they produce supplementary statistical information; in the sense that the contextual probability distribution of, e.g., the observable, $b$ could not be reconstructed on the basis of the probability distribution of $a$. The classical formula of total probability is violated; supplementarity of the observables $a$ and $b$ under the context $C$ induces interference of probabilities $p_C^b(x)$ and $p_C^a(y).$}}

\section{The Hilbert space representation of contexts based on conjugate observables}
 {\bf Definition 6.1 }{\it Observables $a$ and $b$ are said to be symmetrically conditioned if}
\begin{equation}
\label{SC}
p^{a/b}(\alpha/\beta) = p^{b/a}(\beta/\alpha)
\end{equation}

{\bf Definition 6.2 }{\it  Observables $a$ and $b$ are called (statistically)
conjugate if they are symmetrically conditioned and nondegenerate:
 $$p^{a/b}(\alpha/\beta) = p^{b/a}(\beta/\alpha)>0$$
 for all $\alpha, \beta.$}

We shall see that statistically conjugate observables $a$ and $b$ can be represented by noncommutative operators $\hat a$ and $\hat b$ in the Hilbert space. Everywhere below $a$ and $b$ will be (statistically) conjugate observables.
Suppose that, for every $\beta \in X_b,$ the coefficient of supplementarity $ \vert \lambda(b=\beta/a, C)\vert\leq 1.$ In this case we can introduce new statistical
parameters $\theta(b=\beta/a, C)\in [0,2 \pi]$ and represent the
coefficients of statistical disturbance in the trigonometric form:
$\la(b=\beta/a, C)=\cos \theta (b=\beta/a, C).$ Parameters $\theta(b=\beta/a, C)$ are called  {\it{probabilistic phases.}} We remark that in
general there is no geometry behind these phases. By using the
trigonometric representation of the coefficients $\lambda$ we obtain
the well known  {\it formula of interference of probabilities} which
is typically derived by using the Hilbert space formalism.

If both coefficients $\lambda$ are larger than one, we can
represent them as $\la(b=\beta/a, C)=\pm \cosh \theta (b=\beta/a, C)$ and
obtain the formula of hyperbolic interference of probabilities;
there can also be found models with the  mixed
hyper-trigonometric behavior, see Refs. 29, 30.

We consider only
the complex Hilbert space representation of trigonometric contexts:
$$
{\it  C}^{\rm tr}=\{C :|\la(\beta/a, C)|\leq 1,
\beta\in X_b\}.
$$
Of course, the system ${\it  C}^{\rm tr}$ depends on the
choice of a pair of reference observables, ${\it  C}^{\rm tr}\equiv
{\it  C}^{\rm tr}_{b/a}.$ We set $p_C^a(\alpha)={\bf P}(a=\alpha/C),
p_C^b(\beta)={\bf P}(b=\beta/C), p(\beta/\alpha)={\bf P}(b=\beta/a=\alpha), \beta \in X_b, \alpha \in X_a.$
(In previous sections we considered probabilities with respect to
a fixed context $C;$ therefore this index was omitted).
 Let context $C\in {\it  C}^{\rm{tr}}.$ The interference formula of total probability (\ref{KOL11}) can be
written in the following form:
$$
p_c^b(\beta)=\sum_{\alpha \in X_a}p_C^a(\alpha) p(\beta/\alpha)
+ 2\cos
\theta_C(\beta)\sqrt{\Pi_{\alpha \in X_a}p_C^a(\alpha) p(\beta/\alpha)},
$$
where $\theta_C(\beta)=\theta(b=\beta/a, C)= \pm \arccos \lambda(b=\beta/a, C),
\beta \in X_b.$ By using the elementary formula: $D=A+B+2\sqrt{AB}\cos
\theta=\vert \sqrt{A}+e^{i \theta}\sqrt{B}|^2,$ for $A, B > 0,
\theta\in [0,2 \pi],$ we can represent the probability $p_C^b(\beta)$ as
the square of the complex amplitude (Born's rule):
\begin{equation}
\label{Born} p_C^b(\beta)=\vert \psi_C(\beta) \vert^2 ,
\end{equation}
where a complex probability amplitude is defined  by
$$\psi(\beta)\equiv \psi_C(\beta)$$
\begin{equation}
\label{EX1} =\sqrt{p_C^a(\alpha_1)p(\beta/\alpha_1)}
+ e^{i \theta_C(\beta)} \sqrt{p_C^a(\alpha_2)p(\beta/\alpha_2)} \;.
\end{equation}
We denote the space of functions: $\psi: X_b\to {\bf C}$ by the
symbol $\Phi =\Phi(X_b, {\bf C}),$ where ${\bf C}$ is the field of complex numbers.
Since $X_b= \{\beta_1, \beta_2 \},$ the
$\Phi$ is the two dimensional complex linear space. By using the
representation (\ref{EX1}) we construct the map $J^{b/a}:{\it  C}^{\rm{tr}}
\to \Phi(X_b, {\bf C})$ which maps contexts (complexes of, e.g.,
physical conditions) into complex amplitudes. The representation
({\ref{Born}}) of probability is nothing other than the famous {\bf
Born rule.} The complex amplitude $\psi_C(\beta)$ can be called a
{\bf wave function} of the complex of physical conditions (context)
$C$  or a  (pure) {\it state.}  We set $e_\beta^b(\cdot)=\delta(\beta -
\cdot).$ The Born's rule for complex amplitudes (\ref{Born}) can be
rewritten in the following form:
\begin{equation}
\label{BH}
p_C^b(\beta)=\vert(\psi_C, e_\beta^b)\vert^2,
\end{equation}
where the scalar product
in the space $\Phi(X_b,{\bf C})$ of complex amplitudes is defined by the standard formula:
\begin{equation}
\label{RRRR}
(\psi_1, \psi_2) = \sum_{\beta\in X_b} \psi_1(\beta)\bar \psi_2(\beta).
\end{equation}
 The system
of functions $\{e_\beta^b\}_{\beta\in X_b}$ is an orthonormal basis in the
Hilbert space $H=(\Phi, (\cdot, \cdot))$ Let $X_b \subset {\bf R},$ where ${\bf R}$ is the field of real numbers.
By using
the Hilbert space representation  of the Born's rule  we
obtain  the Hilbert space representation of the classical conditional expectation of the
observable $b$:
$$
E (b/C)= \sum_{\beta\in X_b} \beta \;p_C^b(\beta)= \sum_{\beta \in X_b} \beta \;\vert\varphi_C(\beta)\vert^2
$$
$$
= \sum_{\beta\in X_b}\beta \;(\psi_C, e_\beta^b)
\overline{(\psi_C, e_\beta^b)}= (\hat b \psi_C, \psi_C),
$$
where
the  (self-adjoint) operator $\hat b: H \to H$ is determined by its
eigenvectors: $\hat b e_\beta ^b = \beta e^b_\beta, \beta \in X_b.$ This is the
multiplication operator in the space of complex functions
$\Phi(X_b,{\bf C}):$ $ \hat{b} \psi(\beta) = \beta \psi(\beta). $ It is
natural to represent this observable (in the Hilbert space
model)  by the operator $\hat b.$ We would like to have Born's rule
not only for the $b$-observable, but also for the $a$-observable:
$$
p_C^a(\alpha)=\vert(\psi, e_\alpha^a)\vert^2 \;, \alpha \in  X_a .
$$
How can we define the basis $\{e_\alpha^a\}$ corresponding to the
$a$-observable? Such a basis can be found starting with interference
of probabilities. We set $u_j^a=\sqrt{p_C^a(\alpha_j)},
p_{ij}=p(\beta_j/\alpha_i), u_{ij}=\sqrt{p_{ij}}, \theta_j=\theta_C(\beta_j).$ We
have:
\begin{equation}
\label{0} \psi=u_1^a e_1^a + u_2^a e_2^a,
\end{equation}
where
\begin{equation}
\label{Bas} e_1^a= (u_{11}, \; \; u_{12}) ,\; \; e_2^a= (e^{i
\theta_1} u_{21}, \; \; e^{i \theta_2} u_{22})
\end{equation}
We consider the {\it matrix of transition probabilities} ${\bf
P}^{b/a}=(p_{ij}).$ It is always a  {\it stochastic matrix:}
$p_{i1}+p_{i2}=1, i=1,2).$ We remind  that a matrix is called  {\it
double stochastic} if it is stochastic and, moreover, $p_{1j} +
p_{2j}=1, j=1,2.$ The  system $\{e_i^a\}$   is an orthonormal basis iff the matrix ${\bf
P}^{b/a}$ is double stochastic and probabilistic phases satisfy the
constraint: $ \theta_2 - \theta_1= \pi \; \rm{mod} \; 2 \pi,$ see Refs. 29, 30

.

Thus if the matrix ${\bf P}^{b/a}$ is double
stochastic, then the $a$-observable is represented by the operator
$\hat{a}$ which is diagonal (with eigenvalues $\alpha)$ in the basis
$\{e_\alpha^a\}.$ The  classical conditional average of the
observable $a$ coincides with the quantum Hilbert space average:
$$
E(a/C)=\sum_{\alpha \in \alpha} \alpha p_C^a(\alpha) = (\hat{a} \phi_C, \phi_C), \; C \in {\it
C}^{\rm{tr}}.
$$

\section{Complementarity, supplementarity and mutual exclusivity}
I start this section with an extended citation from the report of one of the referees. The problems discussed by those referee are of great importance for clarifying the role of the mutual exclusivity in Bohr's principle of complementarity:

{\small ``I would like now to comment on the author's "supplementarity principle". The author juxtaposes this principle to Bohr's complementarity principle, specifically on the account of the mutual exclusivity of whatever {\it{elements}} are involved in a given complementary situation. (I shall explain the term "elements" and my emphasis presently.) According to Bohr's interpretation, quantum mechanics is characterized by complementary physical situations, say, those of the position and momentum measurements for a given quantum object. Bohr's terminology refers to situations that are mutually exclusive and hence not applicable at the same time, and yet both defined as possible in order to achieve a comprehensive (complete) physical description. This mutual exclusivity is no longer required by the author's supplementarity principle, and it is replaced by the rule for conditioned probabilities for the corresponding observables, such as position and momentum.

It appears to me that there is some misunderstanding or confusion here as concerns Bohr's complementarity principle, which is not uncommon in treatments of Bohr. Indeed, in the present case it may be a matter of further clarification or, again, a greater lucidity of exposition than misunderstanding. I should add that one's understanding of Bohr's complementarity principle may depend on which Bohr one reads; in particular there are differences between his earlier and his later (e.g., post-EPR) views concerning complementarity. To some degree, the nature of the "elements" involved in complementary situations change in Bohr's later view of the quantum-mechanical situation (hence my emphasis above). In Bohr's later view, to which the author refers, the situation may be seen as follows.

Such variables as position and momentum not only cannot be measured but also cannot be assigned or indeed defined simultaneously, as the author suggests. It is important, however, that this argumentation applies strictly to a given physical situation of measurement, whereby such {\it{variables}} as position or momentum (understood in terms of classical physics) and the corresponding measurable physical {\it{properties}} now refer only to measuring instruments impacted by quantum objects.

The author seems to me to be insufficiently attentive to these nuances, especially those concerning the differences between mathematical {\it{variables}} and physical {\it{properties}}, and between each of these and {\it{observables}}. Part of the problem is the ambiguity of the term "observable" in quantum mechanics. It can refer to both an observable property and a corresponding variable, say, as an operator in a Hilbert space. The relationships between both, essentially different, types of "observables", Hilbert-space operators and measurable quantities, and specifically the assignment of probabilities to the outcomes of measurements that result, may be seen as defining the essence of quantum mechanics.

In Bohr, an assignment and the definition of certain physical properties (which, again, pertain to the measuring instruments involved and are seen as classical physical variables) would always be mutually exclusive, and the assignment of probabilities would be affected accordingly. The definition of mathematical variables as Hilbert-space operators is, however, a more complex matter. Both variables potentially involved in one or the other complementary situation of measurement possible at any given point may be viewed as being, at that point, definable {\it{simultaneously}} (rather than being mutually exclusive) as {\it{formal mathematical entities}}. The reason for that is that we consider formally the same Hilbert space in any given situation of measurement, where we would, for example, consider the commutators corresponding to the uncertainty relations for such variables.''}

I agree with the referee that in this paper there is investigated "mathematical mutual exclusivity" and the physical consequences of our principle of supplementarity should be investigated in more detail. We shall do this in section 8 by replying to the questions of another referee of the paper.

Now we discuss mainly mathematical framework. I agree with the referee that one should distinguish between mathematical variables, physical properties and observables. The essence of quantum mechanics (as physical theory) is in understanding relationships between these mathematical and physical quantities. Denote mathematical variables $u$ and $v$, physical properties ${\cal U}$ and ${\cal V}$ and (physical) observables $U$ and $V$. Bohr's mutual exclusivity is mutual exclusivity of $U$ and $V$. Thus by mutual exclusivity N. Bohr understood exclusivity of measurement contexts $D_U$ and $D_V,$ cf., with the remark in introduction on the difference between Bohr's ``measurement contextuality" and our ``preparation contextuality". Physical observables $U$ and $V$ are represented in quantum formalism by self-adjoint operators $\hat u$ and $\hat v.$ As was rightly pointed out by the referee, $\hat u$ and $\hat v$ are definable
simultaneously as formal mathematical quantities. The situation with physical properties ${\cal U}$
and ${\cal V}$ is essentially more complicated. Logically mutual exclusivity of observables $U$ and $V$ does not automatically imply that physical properties ${\cal U}$ and ${\cal V}$ could not coexist. The views of N. Bohr on this problem are not so clear. One could not say that Bohr rejected the existence of physical properties. The result of a measurement was considered by him as a property of the object, see [] for the extended discussion. However, this result can be defined only in the context of his measurement. Therefore for mutually exclusive measurement contexts $D_U$ and $D_V$ it is forbidden to consider the properties ${\cal U}$ and ${\cal V}$ as simultaneously coexisting. It should be pointed out the simultaneous definability of mathematical variables $\hat u$ and $\hat v$ (self-adjoint operators in the complex Hilbert space) does not contradict to mutual exclusivity of the corresponding physical observables. The quantum formalism does not describe results of individual measurements.
Therefore operators $\hat u$ and $\hat v$ could not be assigned to a single physical system $\om.$

 Moreover, it seems that various ``no-go" theorems, e.g., theorems of von Neumann, Kochen-Specker, Bell, confirm Bohr's views to complementarity. These theorems imply that it is even in principle impossible to go deeper that quantum formalism. It is impossible to construct a prequantum (`classical") probabilistic model containing mathematical variables $u$, $v$ which could be assigned to individual systems.\footnote{ I do not know any Bohr's comment on the possibility to justify his principle of complementarity via ``no-go" theorems, in particular, von Neumann theorem. It seems that N. Bohr would not be so interested in such results. For him already Heisenberg's uncertainty relations and the impossibility to combine the position measurement with the interference in the two slit experiment were sufficiently strong arguments in favor of complementarity (in the sense of mutual exclusivity).}

  The V\"axj\"o model induces a totally different viewpoint to relationships between mathematical variables, physical properties and observables. In the opposite to the common ``quantum opinion" in this model it is possible to define mathematical variables $u=a$ and $v=b$ which are simultaneously defined for a single physical system $\om:  a(\om)$ and $b(\om)$ coexist. These functions represent physical observables $U$ and $V$ which were denoted by the same symbols $a$ and $b$. Here objective properties ${\cal U}$ and ${\cal V}$ can coexist and they are represented by $a(\om)$ $b(\om)$.

As was pointed out, the conventional quantum formalism is a special representation of the V\"axj\"o model in which one ignores the knowledge about $a(\om)$ and $b(\om).$ Such an ignorance is convenient in the situation in that the simultaneous measurement of $U=a$ and $V=b$ is impossible. Here impossibility need not be of fundamental nature, it could be of purely technological nature, cf. section 8.

Quantum model can not describe a single-particle context $C_{\omega},$ but the prequantum contextual model can contain such contexts.

As a consequence of this difference of models Bohr's principle of complementarity does not look so natural for the V\"axj\"o model and should be changed to a new principle. We propose to consider supplementarity with coexistence instead complementarity (with mutual exclusivity).

\section{Experimental consequences of the supplementarity principle}

One of the referees of this paper asked about possible experimental
consequences of the supplementarity principle. He pointed out that:
{\small ``According to Bohr, observables $a(\omega), b(\omega),$ when
measured on a system $\omega,$ may be considered as relevant
information about $\omega$ only when they are compatible, while in
Khrennikov's realistic contextual model the contrary is claimed to
be the case. In order to distinguish both approaches, the author
introduces the new notions of supplementarity, and supplementary
observables. Khrennikov's claim may be criticized, as Bohr's
complementarity standpoint and the associated incompatibility of
some specific observables fits perfectly actual laboratory practice.
However, nobody knows how physics and present day technical
possibilities will evolve in the future and, therefore, the possible
theoretical alternative considered by Khrennikov may not be excluded
a priori.''} Then the referee added that the physical relevance and
interest of the principle of supplementarity should be explained
somewhat more explicitly. {\small ``Relevant questions are e.g. how
the principle of supplementarity related to experiment, how may one
profit from additional information? Or is it at present only a
theoretical possibility, which may become interesting in the
future?''}

At the moment the principle of supplementarity has purely
theoretical value. It says that quantum probabilistic behavior need
not imply the impossibility to construct a realistic underlying
model. Therefore one cannot exclude the possibility to find (or
create) a context $C$ such that Heisenberg's uncertainty relations
would be violated in this context for some pair of conjugate
variables $a$ and $b.$ Here Heisenberg's uncertainty relations are
considered as an inequality for dispersions of $a$ and $b$ for a
statistical ensemble $S_C$ corresponding to the context $C,$ see
e.g. L. Ballentine$^{(14)}.$ Thus there might be found a context $C$ such
that dispersions of both $a$ and $b$ would be arbitrary small. Of course, such a context
$C$ would be impossible to represent by a complex probability amplitude $\psi_C.$
Moreover, there could exist  {\it dispersion free} contexts. There is no
contradiction with Neumann's conclusion on nonexistence of   dispersion free contexts
(because in our model such contexts are not represented in the complex Hilbert space).
In general the set of trigonometric contexts ${\cal C}^{\rm{tr}}$ (here $\vert \lambda\vert \leq 1,$ see section 2.9) which can be represented in the complex Hilbert space is a proper subset of the family of all possible contexts  ${\cal C}$ of a contextual statistical model $M =({\cal C}, {\cal O}, {\cal D}({\cal O}, {\cal C})).$ We recall that there also exist hyperbolic contexts, ${\cal C}^{\rm{hyp}}$ (here
$\vert \lambda\vert \geq 1)$ which can be represented in so called hyperbolic Hilbert space.
It is possible to show that  dispersion free contexts belong to the class
${\cal C}\setminus ({\cal C}^{\rm{tr}}\cup {\cal C}^{\rm{hyp}}).$
\footnote{It is easy to present mathematical models$^{(1)}$ in that ${\cal C}\setminus ({\cal C}^{\rm{tr}}\cup {\cal C}^{\rm{hyp}})\not=\emptyset$
and there exist dispersion free contexts. But the referee is
completely right that  until now such contexts have not been found.
There are two possibilities:   either they do not exists in nature
(as it follows from the conventional viewpoint to QM) or there were
never performed extended experimental investigations in this
direction. I hope that the development of quantum cryptography
(which is fundamentally based on Bohr's complementarity principle)
could clarify this problem.}

Another consequence of the principle of supplementarity which can
stimulate experimental research is that quantum-like probabilistic
effects might be observed for systems and contexts (e.g. physical or
biological) for which existence of an underlying realistic model
looks very natural. Here an observation of interference of
probabilities would not be considered as a contradiction with the
realistic model. As was mentioned in Ref.1, such effects might be
observed for ensembles of macroscopic physical systems (related to
specially designed contexts) or for ensembles of cognitive systems,
see the previous section.\footnote{ We remark that in cognitive science it is
commonly believed that there can be created a
realistic neurophysiological model of brain functioning, see e.g. Ref. 47.
It is natural to suppose that such a realistic model would be
of huge complexity. However,  it is not so easy to derive some conclusions
about individual behavior of a system (brain) depending on billions
of (neuronal) parameters. Therefore in such a situation a
probabilistic model plays the important role. If we apply Bohr's the complementarity principle, then it seems  that the quantum
probabilistic model cannot be applied in this case (in the presence of
underlying realistic neurophysiological model). However, if we apply
the contextual model, then it is very natural  to apply
quantum-like model to describe the  probabilistic behavior of brain
(as the whole), see Refs. 40--42. For example, R. Penrose$^{(48)}$  who
evidently uses Bohr's principle of complementarity remarked:
``It is hard to see how one could usefully consider a quantum superposition of one neuron {\it firing} and {\it not firing.''} Therefore he was not able to apply quantum formalism on the neurophysiological
macroscale and, as a consequence, he should go to the deepest level of matter described by quantum gravity.}

The referee also proposed to discuss in more detail relation between
the present contextual statistical realist model and models with
hidden variables (HV). Surprisingly our model does not have close
relation with HV-models. Despite the realism of reference
observables ($a(\omega), b(\omega)$ are well defined for any
physical system $\omega),$ we do not assume the existence of the
simultaneous (frequency) probability distribution (as people do in
all HV-models), but the most important is that even for each single
variable probability distributions are contextual -- determined by
complexes of experimental physical conditions. We did not create
a HV-model for QM. The only thing that was demonstrated is that the interference of probabilities is compatible with the realistic viewpoint to (reference) observables.

\section{Appendix 1: Bell's inequality}

{\small Typically Bell's inequality is considered as
a constraint for local realistic models. The question of relation of Bell's inequality with interpretations of QM was discussed in details by L. Ballentine$^{(14)},$
De Muynck, De Baere and Marten$^{(38)},$  De Muynck$^{(17)}.$ Our contextual statistical realistic interpretation does not contradict to the experimental fact of the violation of Bell's inequality. In our model only {\it two} reference observables are realistic and
Bell's theorem is about realism of {\it three} observables. But it may be that is not the point!
Bell's realism is a very special statistical realism, namely, the Kolmogorovian realism. J. Bell
claimed from the very beginning that there exist the Kolmogorov probability measure
$\rho(d\lambda)$ on the space of hidden variables $\Lambda$ (see, e.g., Accardi$^{(15)},$
see also Ref. 49, 50). But in general there are no reasons for the existence of such a unique
probability measure on $\Lambda.$ For example, let us assume that $\Lambda$
is the infinite dimensional topological linear space (e.g., Hilbert or Banach). This assumption is not so unnatural -- it is clear that the space of hidden variables should be of the greatest complexity$^{(49)}.$ Moreovere, De Muynck, De Baere and Marten$^{(38)}$
and De Muynck$^{(17)}$ proposed to consider quantum observables as averages with respect to
trajectories, see also Ref. 49 (where that model was investigated). Such spaces are, of course, infinite dimensional. It is well know that many natural distributions on infinite
dimensional spaces are not countably additive (for example, some Gaussian measures,
see Ref. 49 for analysis). Such distributions are not probability measures; for them
Bell's calculations inducing the Bell inequality cannot be performed. Recently it was shown
that if, instead of the Bell-Kolmogorov realism, one uses the frequency probabilistic realism
(von Mises realism), then  in general there is no Bell's inequality$^{(49)}$ (neither
GHZ-paradox) and, moreover, it is possible to obtain the EPR-Bohm correlations. These
correlations in the general contextual probabilistic framework were obtained in Ref. 25.
In fact, the using of von Mises realism is closely related with objections to Bell's
arguments presented by W. De Baere$^{(51)}$ (the hypothesis of nonreproducibility) and
K. Hess and W. Philipp$^{(52)}$ (taking into account the time-structure of the experiment).}

\section{Appendix 2: Supplementarity and Kolmogorovness}

{\bf Definition 10.1.} {\it{ Probabilistic data $D(a, b,  C)$
is said to be Kolmogorovian if there exists a Kolmogorov probability space
${\cal P}=(\Omega, {\cal F}, {\bf P})$ and random variables $\xi_a$ and $\xi_b$ on $\cal P$ such that:}}
\begin{equation}
\label{KA}
p^a(\alpha)={\bf P}(\xi_a=\alpha), \; \; p^b(\beta)={\bf P}(\xi_b=\beta);
\end{equation}
\begin{equation}
\label{KA1}
p^{b/a}(\beta/\alpha)={\bf P}(\xi_b=\beta/\xi_a=\alpha), \; \; p^{a/b}(\alpha/\beta)={\bf P}(\xi_a=\alpha/\xi_b=\beta).
\end{equation}

If data $D(a, b,  C)$ is Kolmogorovian then the observables $a$ and $b$
can represented by Kolmogorovian random variables $\xi_a$ and $\xi_b.$

{\bf Lemma 10.1.} {\it Data $D(a, b,  C)$ is Kolmogorovian if and only if}
\begin{equation}
\label{AX}
p^a(\alpha) p^{b/a}(\beta/\alpha)=p^b(\beta) p^{a/b}(\alpha/\beta)
\end{equation}

{\bf Proof.} a) If data $D(a, b,  C)$ is Kolmogorovian then (\ref{AX}) is
reduced to the equality ${\bf P}(O_1 \cap O_2)={\bf P} (O_2 \cap O_1)$ for  $O_1, O_2 \in {\cal F}.$

b) Let (\ref{AX}) holds true. We set $\Omega= X_a \times X_b,
 X_a= \{\alpha_1, \alpha_2\}, X_b=\{\beta_1,
\beta_2\}.$ We define the probability distribution on $\Omega$ by
$$
{\bf P}(\alpha, \beta)=p^b(\beta) p^{a/b}(\alpha/\beta)=p^a(\alpha)
p^{b/a}(\beta/\alpha);
$$
and define the random variables $\xi_a(\omega)=\alpha,
\xi_b(\omega)=\beta$ for a system $\omega$ on which the outcomes
$\alpha, \beta$ are found when observables $a,b$ are measured. We
have

${\bf P}(\xi_a(\omega)=\alpha)=\sum_\beta {\bf P}(\alpha,
\beta)=\sum_\beta p^a(\alpha) p^{b/a}(\beta/\alpha)$

$=p^a(\alpha) \sum_\beta p^{b/a} (\beta/\alpha)=p^a(\alpha).$

And in the same way  ${\bf P}(b=\beta)=p^b(\beta).$ Thus
$$
{\bf P}(a=\alpha/b=\beta)=\frac{{\bf P}(a=\alpha, b=\beta)}{{\bf P}(b=\beta)}=
\frac{p^b(\beta) p^{a/b}(\alpha/\beta)}{p^b(\beta)}=p^{a/b}(\alpha/\beta).
$$
And in the same way we prove that $p^{b/a}(\beta/\alpha)={\bf P}(b=\beta/a=\alpha).$

\medskip

We now investigate the relation between Kolmogorovness and nonsupplementarity.
If $D(a, b,  C)$ is Kolmogorovian then the formula of total probability holds
true and we have (\ref{C3}). Thus observables $a$ and $b$ are nonsupplementary (in the
context $C).$ Thus:

\medskip

{\it Komogorovness implies nonsupplementarity}

or as we also can say:

{\it Supplementarity  implies non-Kolmogorovness.}

\medskip

However, in the general case nonsupplementarity does not imply
that probabilistic data $D(a, b,  C)$ is Kolmogorovian. Let us investigate in more detail
the case when both matrices ${\bf P}^{a/b}$ and ${\bf P}^{b/a}$ are {\it double stochastic.}
We recall that a matrix ${\bf P}^{b/a}= (p^{b/a}(\beta/\alpha))$ is double stochastic if it is
stochastic (so (\ref{SS}) holds true) and, moreover,
\begin{equation}
\label{SSS}
\sum_\alpha p^{b/a}(\beta/\alpha) =1, \beta= \beta_1, \beta_2.
\end{equation}

{\bf Remark 10.1.} (Double stochasticity as the law of statistical balance)
{\small As was mentioned, the equality(\ref{SS}) holds true automatically.
This is a consequence of additivity and normalization by 1 of the probability distribution
of any collective $x^\alpha.$ But the equality (\ref{SSS}) is an additional condition
on the observables $a$ and $b.$ Thus by considering double stochastic matrices
we choose a very special pair of reference observables. In Ref. 2 I tried to find the physical meaning of
the equality (\ref{SS}).  Since $p^{b/a}(\beta/\alpha_2)= 1 - p^{b/a}(\beta/\alpha_1),$
the $C_{\alpha_1}$ and $C_{\alpha_2}$ contexts compensate each other in ``preparation of the
property'' $b=\beta.$  Thus the equation (\ref{SSS}) could be interpreted the {\it law of statistical
balance} for the property $b=\beta.$ If both matrices ${\bf P}^{b/a}$ and ${\bf P}^{a/b}$
are double stochastic then we have laws of statistical balance for both  properties:
$a=\alpha$ and $b=\beta.$}

\medskip

{\bf Definition 10.2.} {\it Observables $a$ and $b$ are said to be statistically balanced
if both matrices ${\bf P}^{b/a}$ and ${\bf P}^{a/b}$ are double stochastic.}

It is useful to recall the following well known result about double stochasticity for Kolmogorovian
random variables:

{\bf Lemma 10.2.} {\it Let $\xi_a$ and $\xi_b$ be random variables on a Kolmogorov space
${\cal P}=(\Omega, {\cal F}, {\bf P})$. Then the following conditions are equivalent

1). The matrices ${\bf P}^{a/b}=({\bf P}(\xi_a=\alpha/\xi_b=\beta) ),\;
{\bf P}^{b/a}=({\bf P}(\xi_b=\beta/\xi_a=\alpha))$ are double stochastic.

2). Random variables are uniformly distributed:

${\bf P}(\xi_a=\alpha) ={\bf P}(\xi_b=\beta)=\frac{1}{2}.$

3). Random variables are symmetrically conditioned:}
\begin{equation}
\label{SIR}
{\bf P}(\xi_a=\alpha/\xi_b=\beta)={\bf P}(\xi_b=\beta/\xi_a=\alpha).
\end{equation}

This result is not true in the contextual frequency approach and this fact is used
in the following proposition:

{\bf Proposition 10.1.} {\it A Kolmogorov model for data $D(a, b,  C)$ need
not exist even in the case of nonsupplementary statistically balanced
observables having the uniform probability distribution.}

{\bf Proof.} Let us consider probabilistic data $D(a, b,  C)$ such
that $p^a(\alpha)=p^b(\beta)=1/2$ and both matrices ${\bf P}^{a/b}$
and ${\bf P}^{b/a}$ are double stochastic. But let
$p^{a/b}(\alpha/\beta) \ne p^{b/a}(\beta/\alpha).$ Then by Lemma 10.1
data $D(a, b,  C)$ is non-Kolmogorovian, but
$$
2 \delta(\alpha/\beta, C)= 1- \sum_\beta p^{a/b}(\alpha/\beta)=0,\;
2 \delta(\beta/\alpha, C)= 1- \sum_\alpha p^{b/a}(\beta/\alpha)=0.
$$

It seems to be that symmetrical conditioning  plays the cruicial
role in these considerations.

{\bf Lemma 10.3.} {\it If observables $a$ and $b$ are symmetrically conditioned, then
they are statistically balanced (so the
matrices  ${\bf P}^{a/b}$ and ${\bf P}^{b/a}$ are double stochastic).}

{\bf Proof.} We have
$$
\sum_\beta p^{a/b}(\alpha/\beta) = \sum_\beta p^{b/a}(\beta/\alpha)= \sum_\beta {\bf P}_{x^\alpha}(b=\beta)=1;
$$
$$
\sum_\alpha p^{b/a}(\beta/\alpha)= \sum_\alpha {\bf P}_{y^\beta}(a=\alpha)=1 .
$$

\medskip
As we have seen in Proposition 10.1, statistically balanced observables need not be symmetrically conditioned.

{\bf Proposition 10.2.} {\it{ Let observables $a$ and $b$ be symmetrically conditioned. Probabilistic
data $D(a, b,  C)$ is Kolmogorovian iff the observables $a$ and $b$ are nonsupplementary in the context $C.$}}

{\bf Proof.} Suppose that $a$ and $b$ are nonsupplementary.
We set
$$
p^{b/a}(1/1)=p^{b/a}(2/2)=p \; \mbox{ and}\;  p^{b/a}(1/2)=p^{b/a}(2/1)=1-p
$$
(we recall that by Lemma 10.3 the matrix ${\bf P}^{b/a}$ is double stochastic). By (\ref{bC3}), (\ref{bC4})
we have
$$
p^a(\alpha_i)=\sum_\beta p^b(\beta) p^{a/b} (\alpha_i/\beta)=
\sum_\beta \sum_\alpha p^a(\alpha) p^{b/a}(\beta/\alpha)p^{a/b}(\alpha_i/\beta)
$$
$$
=\sum_\alpha p^{a}(\alpha) \sum_\beta p^{b/a}(\beta/\alpha) p^{b/a}(\beta/\alpha_i).
$$
Let us consider the case $i=1:$

$
p^{a}(\alpha_1)=p^{a}(\alpha_1)(p^2 + (1-p)^2)+ 2 p^{a}(\alpha_2) p(1-p)= p^{a}(\alpha_1)(1-4p + 4 p^2) + 2 p(1-p).
$

Thus $p^{a}(\alpha_1)=1/2.$ Hence $p^{a}(\alpha_2)=1/2.$ In the same way we get
that $p^{b}(\beta_1)=p^{b}(\beta_2)=1/2.$ Thus the condition (\ref{AX}) holds true and there exist a
Kolmogorov model ${\cal P}=(\Omega, {\cal F}, {\bf P})$ for probabilistic data $D(a, b,  C).$

\medskip

{\bf Conclusion.} {\it In the case of symmetrical conditioning Kolmogorovness is equivalent to nonsupplementarity. }

\section{Appendix 3: Incompatibility, supplementarity and existence of the joint probability distribution}
The notions of incompatible and complementary variables are considered
as synonymous in the Copenhagen quantum mechanics.
Moreover, compatibility (and consequently noncomplementarity)
is considered as equivalent to existence of the simultaneous probability distribution
$$
{\bf P}(\alpha, \beta) = {\bf P}(a=\alpha, b=\beta).
$$
We now consider similar questions for our
V\"axj\"o model.

{\bf 11.1. Existence of the simultaneous probability distribution.}
By analogy with Definition 5.1 we may propose the following
definition:

{\bf Definition 11. 1.} {\it{ Probabilistic data
$$
W(a, b, C) = \{p^a(\alpha), p^b(\beta)\},
$$
where $C$  is a context, is said to be Kolmogorovian if there exists a Kolmogorov probability space
${\cal P}=(\Omega, {\cal F}, {\bf P})$ and random variables $\xi_a$ and $\xi_b$ on $\cal P$ such that:}}
\begin{equation}
\label{KAw}
p^a(\alpha)={\bf P}(\xi_a=\alpha), p^b(\beta)={\bf P}(\xi_b=\beta).
\end{equation}

However, it is evident that data $W(a, b, C)$ is always Kolmogorovian. For example, we can always define the Kolmogorov
measure
\begin{equation}
\label{KLP}
{\bf P}(\alpha, \beta) = p^a(\alpha) p^b(\beta).
\end{equation}
It is evident that the marginal distributions of this probability coincide with $p^a(\alpha)$ and
$p^b(\beta).$ Of course, such a probability is not uniquely defined and in general this is a purely
mathematical construction which has no physical meaning. In particular, the probability (\ref{KLP})
always corresponds to independent random variables $\xi_a$ and $\xi_b.$ However, in general observables
$a$ and $b$ are not independent. Therefore the problem of Kolmogorovness of data  $W(a, b, C)$ can
be investigated in physical framework only by using the  frequency probability theory.

We emphasize that mathematical Kolmogorovness of data $W(a, b, C)$ does not imply Kolmogorovness of
data $D(a, b, C).$ If data $W(a, b, C)$ is Kolmogorovian, then there are well defined
Bayes-Kolmogorov conditional probabilities:
$$
{\bf P} (A_\alpha/B_\beta) =\frac{ {\bf P} (A_\alpha \cap B_\beta)}{{\bf P} (B_\beta)},
$$
where $A_\alpha= \{\omega \in \Omega: a(\omega)=\alpha\}, B_\beta=\{\omega \in \Omega:
b(\omega)= \beta\}.$ However, in general these measure-theoretical conditional probabilities
do not coincide with experimental conditional probabilities determined in the frequency framework\footnote{But even
in the Kolmogorov framework we cannot define conditional probabilities in the unique way, because
a Kolmogorov measure for data $W(a, b, C)$  is not uniquely defined.}:
$$
p^{a/b}(\alpha/\beta)= {\bf P}_{y^\beta}(\alpha).
$$
In particular, Kolmogorovness of data $W(a, b, C)$ does not imply that probabilistic measures
of supplementarity $\delta(\alpha/\beta, C)$ and $\delta(\beta/\alpha,C)$ are equal to zero.

\medskip

{\bf Conclusion.} {\it Kolmogorovness of data  $W(a, b, C)$ does not imply nonsupplementarity of
observables $a$ and $b$ in the context $C.$}

{\bf 11.2. Incompatible and supplementary observables.}
As usual,  observables $a$ and $b$ are called {\it compatible} in a context $C$
if it is possible to perform a simultaneous observation  of them under $C.$
For any instant of time $t,$ there can be observed a pair of values $z(t)= (a(t), b(t)).$
There is well defined a sequence of results of observations:
$$
z(a,b/C)
= (z_1,z_2,..., z_N,...), \; z_j=(y_j, x_j),
$$
where $y_j=\alpha_1$ or $\alpha_2$ and $x_j=\beta_1$ or $\beta_2.$
Observables $a$ and $b$ are incompatible in a context $C$
if it is impossible to perform a simultaneous observation  of them under $C.$

Observables $a$ and $b$ are said to be {\it statistically compatible} in a context $C$
if they are compatible in $C$ and the $z(a,b/C)$ is an $S$-sequence (or collective).
\footnote{In the opposite case
observables are statistically incompatible.}
Thus there exists the frequency simultaneous probability distribution:
\begin{equation}
\label{FRT}
{\bf P}(\alpha, \beta)\equiv {\bf P}_z (\alpha, \beta) = \lim_{N\to \infty} \frac{n_N(\alpha, \beta; z)}{N}.
\end{equation}
We remark that in general existence of this frequency probability distribution (\ref{FRT}) has nothing
to do with existence of a formal Kolmogorov probability distribution -- Kolmogorovness of data $W(a, b, C).$
As was mentioned, data $W(a, b, C)$ is always Kolmogorovian, but  sequences of results of observations
$y$ and $x$ are not always combinable.

If  $a$ and $b$ are statistically compatible in a context $C$ then data $D(a, b,  C)$ is Kolmogorovian,
because $p^{a/b}(\alpha/\beta)= {\bf P}_z(a=\alpha/b=\beta)$ and $p^{b/a}(\beta/\alpha)=
{\bf P}_z(b=\beta/a=\alpha).$

Thus we have:

\medskip

{\it Statistical compatibility implies Kolmogorovness of data $D(a, b,  C)$}

or

{\it Non-Kolmogorovness of data $D(a, b,  C)$ implies statistical incompatibility}

\medskip

Thus

\medskip

{\it Statistical compatibility implies nonsupplementarity}

or

{\it Supplementarity implies statistical incompatibility}

However, since in general nonsupplementarity does not imply Kolmogorovness of data
$D(a, b,  C),$ see Proposition 5.1, we have:

\medskip

{\it Nonsupplementarity does not imply statistical  compatibility}

\medskip

Thus there can exist a context $C$ such that observables $a$ and $b$ do not produce supplementary information\footnote{Hence all coefficients $\delta(\alpha/b, C), \delta(\beta/a,C)$ are equal to zero.}, but
they do not have the frequency joint probability distribution.
The notions of statistical compatibility and nonsupplementarity are not equivalent.
Hence, the notions of supplementarity and incompatibility are neither equivalent:

\medskip

{\it Statistical incompatibility does not imply supplementarity}

\medskip
Thus there can exist a context $C$ such that observables $a$ and $b$ do not have the frequency joint probability distribution, but at the same time
they do not produce supplementary information.

Let us consider a contextual statistical model $M= ({\cal C}, {\cal O}, {\cal D}({\cal O},{\cal C}).$
If physical observables $a$ and $b$ are (statistically) compatible  for any
$C \in {\cal C}$ then they are called (statistically) compatible in the model $M.$ They are called (statistically)
incompatible in the model $M$ if there exists $C \in {\cal C}$ such that they are (statistically) incompatible for  $C.$

{\bf 11.3. Compatibility does not imply statistical compatibility.} In quantum physics compatibility
of observables -- the possibility to perform a simultaneous observation-- is typically identified
with statistical compatibility -- existence of the frequency simultaneous probability distribution (\ref{FRT}).
This is a natural consequence of the Kolmogorovian psychology. In the frequency probability
theory we should distinguish
compatibility and  statistical compatibility. We present an example in which
observables are compatible, but the limit (\ref{FRT}) does not exist, so observables are
not statistically compatible.
Of course, this means that $y=y(a/C)$ and $x=x(b/C)$ are not combinable.  We need some
well known results about
the generalized probability given by the density of natural numbers,
see Ref. 18 (we recall that  A. N. Kolmogorov
considered the density of natural numbers as an example of probability, but it was before he proposed
the conventional axiomatics). For a subset
$A \subset {\bf N}$ the quantity
$$
{\bf P}(A)=\lim_{N \to \infty} \frac{|A \cap \{1,...,N \}|}{N},
$$
is called the {\it density} of $A$ if the limit exists.
Here the symbol $\vert V \vert$ is used to denote the
number of elements in a finite set $V.$

Let ${\cal G}$ denote the collection of all subsets of
${\bf N}$ which admit density.
It is evident that each finite $A \subset {\bf N}$ belongs
to ${\cal G}$ and ${\bf P} (A)=0.$ It is also evident that
each subset $B = {\bf N} \setminus A,$ where $A$ is finite,
belongs to ${\cal G}$ and ${\bf P} (B)=1$
(in particular,  ${\bf P}({\bf N})=1).$ The reader can easily find examples
of sets $A\in {\cal G}$ such that $0 < {\bf P}(A) <1.$ The ``generalized probability''
${\bf P}$ has the following properties (cf. S. Gudder$^{(11)}$):

{\bf Proposition 11.1.} {\it  Let  $A_1, A_2 \in {\cal G}$ and
$A_1 \cap A_2=\emptyset.$ Then $A_1 \cup A_2 \in {\cal G}$ and}
$$
{\bf P}(A_1 \cup A_2)= {\bf P}(A_1) + {\bf P}(A_2).
$$

{\bf Proposition 11.2.} {\it Let  $A_1, A_2 \in {\cal G}.$  The following
conditions are equivalent:
$$
\begin{array}{rl}
&
1) A_1 \cup A_2 \in {\cal G};\; \;\;
2) A_1\cap A_2 \in {\cal G};
\\ & \\
&
3) A_1 \setminus A_2 \in {\cal G};\; \; \;

4) A_2 \setminus A_1 \in {\cal G}.
\\
\end{array}
$$
There are standard formulas:}
$$
{\bf P}(A_1 \cup A_2)={\bf P}(A_1) + {\bf P}(A_2) - {\bf P}(A_1 \cap A_2);
$$
$$
{\bf P}(A_1 \setminus A_2) = {\bf P}(A_1) - {\bf P}(A_1 \cap A_2).
$$

It is possible to find sets $A, B \in {\cal G}$
such that, for example, $A\cap B \not \in {\cal G}.$
Let $A$ be the set of even numbers.
Take any subset $C \subset A$ which has
no density. In fact, you can find $C$
such that
$$
\frac{1}{N}|C \cap \{1,2,\cdots, N \}|
$$
is oscillating. There happen two cases:
$C \cap \{2n \}=\{ 2n \}$ or $=\emptyset$.
Set
$$
B=C \cup \{2n-1: C \cap \{2n \} = \emptyset\}
$$
Then, both $A$ and $B$ have densities one half.
But $A \cap B =C$ has no density. Thus ${\cal G}$ is not a set
algebra.

We now consider a context $C$ which produces natural numbers. We introduce
two dichotomous observables:
$$
a(n)= I_A(n), \; b(n)= I_B(n),
$$
where $I_O(x)$ is the characteristic function of a set $O.$ We  assume that
these observables are compatible: we can, e.g., look at a number $n$ and find both
values $a(n)$ and $b(n).$\footnote{If reader like he can consider natural numbers
as systems and interpret $a(n)$ and $b(n)$ as values of observables $a$ and $b$
on the system $n.$}
We obtain two $S$-sequences:
$$
y=y(a/C)= (y_1,..., y_N,...), \; x=x(b/C)= (x_1,..., x_N,...), y_j, x_j=0,1.
$$
The frequency probability distributions are well defined:
$$
p^a(\alpha)\equiv {\bf P}_y(\alpha)= 1/2, \;
p^b(\beta)\equiv {\bf P}_x(\beta)= 1/2.
$$
However, the $S$-sequences $y$ and $x$ are not combinable. Thus
observables $a$ and $b$ are not statistically compatible; for example,
the frequency probability ${\bf P}(1, 1)$ does not exist.\footnote{But, of course, there exist
various Kolmogorov probability measures ${\bf P}_{\rm{Kol}}(\alpha, \beta)$ which
have marginal distributions
$p^a(\alpha), p^b(\beta).$}

\medskip

I would like to thank L. Ballentine, P. Busch,  S. Gudder, B. Coecke, G. Mackey, F. Schroeck, E. Prugovecki, E. Beltrametti, G. Cassinelli, M. Ohya, S. Albeverio, O. Smolyanov, G. Jaeger. P. Kwait, A. Leggett,
W. De Baere, W. De Muynck, L. Accardi, I. Volovich, A. Holevo, V. Andreev, E. Loubenets, V. Belavkin, R. Hudson,
A. Aspect, A. Plotnitsky, B. Hiley, C. Fuchs, D. Mermin, D. Greenberger, K. Hess, W. Philipp, H. Rauch, G. ´t Hooft, N\'an\'asiov\'a, S. Pulmanova for discussions on the statistical structure of quantum theory. I also would like to thank
both referees of the paper for their extended comments which were discussed in sections 7 and 8.
This paper was partly supported by EU-Network
"QP and Applications'' and Nat. Sc. Found., grant N PHY99-07949 at KITP, Santa-Barbara.

\medskip

{\bf REFERENCES}

1. N. Bohr, {\it The philosophical writings of Niels
Bohr}, 3 vols. (Woodbridge, Conn.: Ox Bow Press, 1987).

2. D. Hilbert, J. von Neumann, L. Nordheim, {\it Math. Ann.} {\bf 98}, 1 (1927).

3. P. A. M.  Dirac,  {\it The Principles of Quantum Mechanics}
(Oxford Univ. Press, 1930).

4. W. Heisenberg,  {\it Physical principles of quantum theory}
(Chicago Univ. Press, Chicago, 1930).

5. J. von Neumann, {\it Mathematische Grundlagen der Quantenmechanik}
(Springer, Berlin, 1932).

6. J. von Neumann,  {\it  Mathematical foundations
of quantum mechanics} (Princeton Univ. Press, Princeton, N.J., 1955).

7. A. Einstein, B. Podolsky, N. Rosen,  {\it Phys. Rev.} {\bf 47},  777 (1935).

8. N. Bohr,  {\it Phys. Rev.}  {\bf 48}, 696 (1935).

9. A. S. Wightman,  {\it Proc. Symposia in Pure Math.} {\bf 28}, 147 (1976).

10. G. W. Mackey, {\it Mathematical foundations of quantum mechanics}
(W. A. Benjamin Inc., New York, 1963).

11. A. Land\'e, {\it Foundations of quantum theory} (Yale Univ. Press, 1955);
{\it New foundations of quantum mechanics} (Cambridge Univ. Press, Cambridge, 1968).

12. L. De Broglie, {\it The current interpretation of wave mechanics. A Critical Study} (Elsevier, London,
1964).

D. Bohm, {\it Quantum theory} (Prentice-Hall,
Englewood Cliffs, New-Jersey, 1951).

13. S. P. Gudder, {\it Trans. AMS} {\bf 119},  428 (1965);
{\it Axiomatic quantum mechanics and generalized probability theory}
(Academic Press, New York, 1970); ``An approach to quantum probability,''
{\it Quantum Prob. White Noise Anal.} {\bf  13}, 147 (2001).

14. L. E. Ballentine,  {\it Rev. Mod. Phys.} {\bf 42},  358 (1970);
{\it Quantum mechanics} (Englewood Cliffs,
New Jersey, 1989); ``Interpretations of probability and quantum theory,''
in {\it Foundations of Probability and Physics},
A. Yu. Khrennikov, ed, {\it Q. Prob. White Noise Anal.} {\bf  13},  71 (2001);
{\it Quantum mechanics} (WSP,  Singapore, 1998).

15.  L. Accardi, {\it Phys. Rep.} {\bf 77}, 169(1981);
``The probabilistic roots of the quantum mechanical paradoxes,''
in {\it The wave--particle dualism:  A tribute to Louis de Broglie on his 90th
Birthday}, S. Diner, D. Fargue, G. Lochak, and F. Selleri, eds.
(D. Reidel Publ. Company, Dordrecht,  1984), pp. 47-55;
{\it Urne e Camaleoni: Dialogo sulla realta,
le leggi del caso e la teoria quantistica} (Il Saggiatore, Rome, 1997); ``Locality
and Bell's inequality'', in {\it Foundations of Probability and Physics},
A. Yu. Khrennikov, ed,{\it Q. Prob. White Noise Anal.} {\bf  13},1 (2001).

16. P. Mittelstaedt, E. W. Stachow, {\it Int. J. Theor. Phys.} {\bf 22}, 517 (1983);
P. Mittelstaedt, {\it The interpretation of quantum mechanics and the measurement process} (Cambridge Univ. Press, Cambridge, 1997); {\it Ann. Phys.} {\bf 7},  710 (1998);
``Quantum mechanics without probabilities", in {\it John von Neumann and the foundations of quantum physics,}
M. Rédei, M. Stöltzner, eds. (Kluwer Academic, Dordrecht, Holland, 2001), pp. 121-134;
 "Universality and consistency in quantum mechanics - New problems of an old theory", in {\it Física cuántica y realidad. Quantum physics and reality} (Madrid, 2000),  C. Mataix, A. Rivadulla eds. (Editorial Complutense, Madrid, 2002), pp. 197-213.

17.   W. M. De Muynck, ``Interpretations of quantum mechanics,
and interpretations of violations of Bell's inequality'', in {\it Foundations of Probability and Physics},
A. Yu. Khrennikov, ed,
{\it Q. Prob. White Noise Anal.} {\bf  13},  95 (2001);

18.   W. M. De Muynck, {\it Foundations of quantum mechanics, an empiricists approach} (Kluwer, Dordrecht,
2002).

19. A. Yu. Khrennikov, ed.,  ``Foundations of Probability and Physics'',
{\it Q. Prob. White Noise Anal.,}  {\bf 13} ( WSP, Singapore, 2001).

20. A. Yu. Khrennikov, ed., `` Quantum Theory: Reconsideration
of Foundations,'' {\it Ser. Math. Modeling} {\bf 2} (V\"axj\"o Univ. Press, V\"axj\"o, 2002).

21. A. Yu. Khrennikov, ed.,   ``Foundations of Probability and Physics -2,''
{\it Ser. Math. Modeling} {\bf 5} (V\"axj\"o Univ. Press, V\"axj\"o, 2003).

22. A. Yu. Khrennikov, ed.,   ``Quantum Theory: Reconsideration
of Foundations -2,'' {\it Ser. Math. Modeling} {\bf 10} (V\"axj\"o Univ. Press, V\"axj\"o, 2004).

23. A. E. Allahverdyan, R. Balian, T. M. Nieuwenhuizen, in
{\it Foundations of Probability and Physics-3,}  A. Yu. Khrennikov,
ed., AIP Conference Proceedings, 2005, pp. 16-24.

24. O. N\'an\'asiov\'a, {\it Int. J. Theor. Phys.} {\bf 42},
1889(2003).

25.  A. Plotnitsky, ``Quantum atomicity and
quantum information: Bohr, Heisenberg, and quantum mechanics as an
information theory'', in {\it Quantum theory:
reconsideration of foundations},  A. Yu. Khrennikov,ed.( V\"axj\"o Univ. Press,  2002),
pp. 309-343.

26.  A. Plotnitsky, ``Reading Bohr:
Complementarity, Epistemology, Entanglement, and Decoherence,'' in
{\it NATO Workshop Decoherence and its Implications for Quantum
Computations}, A. Gonis and P. Turchi,Eds. (IOS Press,
Amsterdam, 2001),  pp.3--37.

27. A. Plotnitsky, {\it The knowable and unknowable (Modern science, nonclassical thought, and the ``two cultures''} (Univ. Michigan Press, 2002); {\it Found. Phys.}, {\bf 33}, 1649(2003).

28. S. Gudder, S. Pulmannova, {\it Comm. Math. Univ. Carolinae}, {\bf 39}, 645 (1998).

29. A. Yu. Khrennikov,``On foundations of quantum theory,'' in {\it Quantum Theory: Reconsideration of Foundations},  A. Yu. Khrennikov, ed., Ser. Math. Modeling, {\bf 2} (V\"axj\"o University Press, 2002), pp. 163-173.

30. A. Yu. Khrennikov,  {\it Annalen  der Physik} {\bf 12}, 575 (2003);
``On the classical limit for the hyperbolic quantum mechanics,''
quant-ph/0401035; {\it J. Phys.A: Math. Gen.} {\bf 34},  9965 (2001);
{\it Il Nuovo Cimento} B {\bf 117},   267 (2002);
{\it J. Math. Phys.} {\bf 43} 789 (2002);
Ibid {\bf 45}, 902 (2004).

31. K. Svozil, ``Counterfactuals and contextuality,'' in {\it  Quantum Theory: Reconsideration
of Foundations-3,} A. Yu. Khrennikov,
ed., AIP Conference Proceedings, 2005, pp. 425-437.

32. H. Atmanspacher, H. Primas,  ``Epistemic and ontic quantum realities'', in
{\it Foundations of Probability and Physics}-3, A. Yu. Khrennikov,
ed., AIP Conference Proceedings, 2005, pp. 10-25.

33. E. Beltrametti, G. Cassinelli, {\it The Logic of Quantum Mechanics}
(Addison-Wesley Publ. Com., London-Amsterdam, 1981).

34. P. Busch , M. Grabowski, P. Lahti, {\it Operational Quantum Physics}
(Springer Verlag, Berlin, 1995).

35.    E. B. Davies, {\it Quantum Theory of Open Systems} (Academic Press,
London, 1976).

36.   A. S. Holevo, {\it Probabilistic and
Statistical Aspects of Quantum
Theory} (North-Holland, Amsterdam,  1982).

37. Gudder S. P., {\it J. Math. Phys.}, {\bf 11} 431 (1970) ;
{\it Quantum Probability} (Academic Press, San Diego, 1988).

38. W. de Muynck, W. De Baere, H. Martens, {\it Found. of Physics} {\bf 24}, 1589 (1994).

39. A. N. Shiryayev, {\it Probability} (Springer, New York-Berlin-Heidelberg, 1984).

40. A. Yu. Khrennikov, ``On quantum-like probabilistic structure of mental information,''
{\it Open Systems and Information Dynamics} {\bf 11} (3), 267 (2004).

41. E. Conte, O. Todarello,  A. Federici, F. Vitiello, M. Lopane,  A.
Yu. Khrennikov, ``A preliminary evidence of quantum-like behavior in
measurements of mental states,'' in {\it Quantum Theory:
Reconsideration of Foundations-2,} A. Yu. Khrennikov, ed., {\bf 10} (V\"axj\"o
Univ. Press, V\"axj\"o, 2004), pp. 679-702.

A. A. Grib, A. Yu. Khrennikov, K. Starkov, ``Probability amplitude in quantum-like games'', in
{\it Quantum Theory: Reconsideration of Foundations-2}, A. Yu. Khrennikov, ed., , {\bf 10}
(V\"axj\"o Univ. Press, V\"axj\"o, 2004), pp. 703-722.

42. A. Yu. Khrennikov, {\it Information dynamics in cognitive, psychological and
anomalous phenomena} (Kluwer, Dordreht, 2004).

43. R.  von Mises, {\it The mathematical theory of probability and
 statistics} (Academic, London,  1964).

44. A. Yu. Khrennikov, {\it Interpretations of probability} (VSP Int. Sc. Publ.,
Utrecht, 1999).

45. A. N. Kolmogoroff, {\it Grundbegriffe der Wahrscheinlichkeitsrechnung}
(Springer Verlag, Berlin, 1933); reprinted:
{\it Foundations of the Probability Theory}
(Chelsea Publ. Comp., New York, 1956).

46. R. T. Cox, {\it The algebra of probable inference} (J. Hopkins Univ. Press, Baltimore MD, 1961).

47. R. Ashby, {\it Design of mind} (Chapman--Hall, London, 1952);
D. Amit. {\it Modeling brain function} (Cambridge Univ, Press, Cambridge, 1989).

48. R. Penrose, {\it The emperor's new mind} (Oxford Univ. Press, New-York, 1989);
{\it Shadows of the mind}  (Oxford Univ. Press, Oxford, 1994).

49. A. Yu. Khrennikov, {\it Found. Phys.} {\bf 32}, 1159 (2002); {\it Phys. Lett.},
A, {\bf 278}, 307 (2001);  {\it Il Nuovo Cimento} {\bf 119} (2004).

50.  A. Yu. Khrennikov, {\it Found. Phys. Lett.} {\bf 17}, 691 (2004).

51.  W. De Baere,  {\it Lett. Nuovo Cimento} {\bf 39}, 234 (1984);
{\bf 40}, 488(1984); {\it Advances in electronics and electron physics} {\bf 68},
245 (1986).

52. K. Hess and W. Philipp, {\it Proc. Nat. Acad. Sc.} {\bf 98}, 14224 (2001);
{\bf 98}, 14227(2001); {\bf 101}, 1799 (2004); {\it Europhys. Lett.} {\bf 57}, 775 (2002).

\end{document}